\documentclass[aps,prc,showpacs,twocolumn,floatfix]{revtex4-1}
\usepackage[latin1]{inputenc}
\usepackage{bm}
\usepackage{epsfig}
\usepackage{amsthm}
\usepackage{dcolumn}
\usepackage{graphicx}
\usepackage{amsfonts}
\usepackage{float}
\usepackage{amsmath,amssymb}
\def\bit{\begin{itemize}}
\def\eit{\end{itemize}}
\def\bnu{\begin{enumerate}}
\def\enu{\end{enumerate}}
\def\Agoinfty {\:\raisebox{-1.2ex}{$\stackrel{\textstyle\longrightarrow}
{\mbox {\tiny $A\rightarrow\infty$}}$}\:}
\def\etal{{\it et al.}}
\def\go{\rightarrow  }
\def\be{\begin{equation}}
\def\ee{\end{equation}}
\def\br{\begin{eqnarray}}
\def\er{\end{eqnarray}}
\def\brn{\begin{eqnarray*}}
\def\ern{\end{eqnarray*}}

\def\M {{{\cal M}}}

\def\A {{{\cal A}}}
\def\B {{{\cal B}}}

\def\fot{\frac{1}{2}}
\def\bit{\begin{itemize}}
\def\eit{\end{itemize}}
\def\Ket#1{||#1 \rangle}
\def\Bra#1{\langle #1||}
\def\ie{{\it i.e., }}

\def\nn{\nonumber }
\def\bra#1{\langle #1|}
\def\ket#1{|#1 \rangle}
\def\rf#1{{(\ref{#1})}}
\def\ov#1#2{\langle #1 | #2  \rangle }
\def\sixj#1#2#3#4#5#6{\left\{\negthinspace\begin{array}{ccc}
#1&#2&#3\\#4&#5&#6\end{array}\right\}}

\def\go{\rightarrow  }

\def\x{\times}
\def\pb {{\bf p}}
\def\fot{\frac{1}{2}}

\def\mbt{\mbox{\boldmath$\tau$}}
\def\bin#1#2{\left(\negthinspace\begin{array}{c}#1\\#2\end{array}\right)}

\def\rf#1{{(\ref{#1})}}
\def\ov#1#2{\langle #1 | #2  \rangle }
\def\bra#1{\langle #1|}
\def\ket#1{|#1 \rangle}
\def\Ket#1{||#1 \rangle}
\def\Bra#1{\langle #1||}
\def\rb {{\bf r}}
\def\pb {{\bf p}}

\def\be{\begin{equation}}
\def\ee{\end{equation}}
\def\br{\begin{eqnarray}}
\def\er{\end{eqnarray}}

\def\lra{\leftrightarrow}
\def\Mass{\mathrm{M}}

\def\lra{\leftrightarrow}
\def\x{\times}
\def\xb {{\bf r}}
\def\rh {\hat{\bf r}}
\def\ph {\hat{\bf p}}

\def\xh {\hat{\bf x}}
\def\yh {\hat{\bf y}}
\def\pb {{\bf k}}
\def\pb {{\bf p}}
\def\xb {{\bf x}}
\def\yb {{\bf y}}

\def\nn{\nonumber }

\def\go{\rightarrow  }

\def\psip{^\uparrow\psi}
\def\psin{^\downarrow\psi}
\def\Psip{^\uparrow\Psi}
\def\Psin{^\downarrow\Psi}

\def\threej#1#2#3#4#5#6{\left(\negthinspace\begin{array}{ccc}
#1&#2&#3\\#4&#5&#6\end{array}\right)}
\def\sixj#1#2#3#4#5#6{\left\{\negthinspace\begin{array}{ccc}
#1&#2&#3\\#4&#5&#6\end{array}\right\}}

\newcommand{\sDelta}{\mathsf{\Delta}}
\newcommand{\sgn}{\mathrm{sgn}}
\begin{document}
\title{Relativistic model for the nonmesonic weak decay of single-lambda  hypernuclei}
\author{C.E. Fontoura$^1$}
\author{F. Krmpoti\'c$^{1,2}$}
\author{A.P. Gale\~ao$^{1}$}
\author{C. De Conti$^{3}$}
\author{G. Krein$^{1}$}
\affiliation{$^1$Instituto de F\'isica Te\'orica, Universidade Estadual Paulista \\
Rua Dr. Bento Teobaldo Ferraz, 271 - Bloco II, 01140-070, S\~ao Paulo, SP, Brazil}
\affiliation{$^2$Instituto de F\'isica La Plata, Universidad Nacional de La Plata, 
1900 La Plata, Argentina}
\affiliation{$^3$Campus Experimental de Rosana, Universidade Estadual Paulista, 
19274-000 Rosana, SP, Brazil}
\date{\today}
\begin{abstract}
Having in mind its future extension for theoretical investigations related to charmed nuclei, we develop a relativistic 
formalism for the nonmesonic weak decay of single-$\Lambda$ hypernuclei in the framework of the independent-particle shell model and with the dynamics represented by the $(\pi,K)$ one-meson-exchange model. Numerical results for the one-nucleon-induced transition rates of ${}^{12}_{\Lambda}\textrm{C}$ are presented and compared with those obtained in the analogous nonrelativistic calculation. 
There is satisfactory agreement between the two approaches, and the most noteworthy difference is that the ratio $\Gamma_{n}/\Gamma_{p}$ is appreciably higher and closer to the experimental value in the relativistic calculation. Large discrepancies between ours and previous relativistic
calculations are found, for which we do not encounter any fully satisfactory  explanation. The most recent experimental data is well reproduced 
by our results. In summary, we have achieved our purpose to develop a reliable model for the relativistic calculation of the nonmesonic weak decay of 
$\Lambda$-hypernuclei, which can now be extended to evaluate similar processes in charmed nuclei.
\end{abstract}
\maketitle
\section{Introduction}
\label{sec:intro}
Investigations of exotic nuclear properties, such as large isospin (manifest in the 
so called neutron-rich isotopes), or non\-trivial values of flavor quantum numbers (strangeness, 
charm or beauty), are of continuous interest. The best known nuclei within the last category are 
those where a $\Lambda$-hyperon, with strangeness ${\sf S}=-1$, replaces one of the
nucleons, giving to the composed system some quite unusual pro\-perties.
Such nuclei are referred to as $\Lambda$-hypernuclei {\textemdash} for recent reviews, see Refs.~\cite{Bo12,{Feliciello:2014ola}}.

One of the most remarkable properties of $\Lambda$-hyper\-nuclei is the occurrence of
the nonmesonic weak decay (NMWD), induced by the elementary process 
\be
\Lambda+N\go n+N,
\label{1.1}
\ee
with $N = p \, ({\rm proton})$ or $n \, ({\rm neutron})$. This is the main decay channel for medium- 
and heavy-weight hypernuclei {\textemdash} Refs.~\cite{Garbarino:2013rwa}~and~\cite{Bufalino:2013qwa} 
provide, respectively, reviews on recent theoretical and experimental developments in the study of 
hypernuclear decay; for earlier comprehensive reviews on theory see 
Refs.~\cite{OsetRamos,AlbericoGarbarino,Parreno} and on experiment 
Refs.~\cite{Outa,Pocho}. NMWD can only take place within the nuclear environment 
and is a unique opportunity offered by nature to access the 
strangeness-changing interaction 
between baryons. Its~mean lifetime has been measured in several $\Lambda$-hypernuclei and found
to be of the same order of magnitude as the full mean lifetime of $\Lambda$ 
in free space, $\tau_\Lambda = (2.632 \pm 0.020)\times 10^{-10}\,{\rm s}$~\cite{PDG}.

The NMWD dynamics is frequently handled by one-meson-exchange (OME) models. Such models are
motivated by the fact that the $NN$ in\-teraction at long distance is due to the 
one-pion-exchange, but with the difference that in NMWD the exchange processes occur with one strong and one weak vertex and can include other mesons in addition to the~$\pi$, 
like the pseudoscalar ($K, \eta$) and vector ($\rho,\omega,K^*$) mesons\cite{Du96,Pa97,It02,It03,It08,Ba02,Kr03,Ba03,Ba05,Ba07,Ba08,Ba10,Kr10,Kr10a,Go11,Kr14,Kr14a}. The coupling constants at the strong vertices 
can be taken from different OME models for the $NN$ interaction, while those at the weak vertices
can be extracted from free $\Lambda$ decay data and making use of soft meson theorems and 
$SU(6)_W$ symmetry~\cite{Du96,Pa97}. A recent study conducted within a nonrelativistic 
framework~\cite{Kr14} indicates that $\pi$ and $K$ exchanges give the main contributions to the 
NMWD of $s$-shell hypernuclei.

Instead of implanting a $\Lambda$ in a nucleus one could also imagine to implant a charmed baryon,
like e.g. a $\Lambda^+_c$, in view of the similarity between the quark structures of the strange 
and charmed hyperons, namely $\Lambda(uds)$ and $\Lambda^+_c(udc)$. Such a possibility was in fact 
conjectured 40~years ago~\cite{Tyap} and several authors in the succeeding decades have found, 
using different models for the interactions between nucleons and charmed hyperons, that such
hypothetical exotic nuclei (including even bottom nuclei) could actually form a rich spectrum of 
bound states over a wide range of atomic numbers~\cite{Iwao,Do77,GatPac,Kol,Bham,Bando,
Gib,StaTsa,Cai,Tsu1,Tsu2,Tsu3}. Like $\Lambda$-hypernuclei, $\Lambda^+_c$-hypernuclei may also decay via a NMWD process. One example is~\cite{Bu92}   
\be
\Lambda^+_c + n \go \Lambda + p,
\label{1.2}
\ee
which can be induced by the exchange of a $\pi,\rho$ or $K$ meson. Another possibility is 
\be
\Lambda^+_c + N \go p + N,
\label{1.3}
\ee
induced by the exchange of a $D$ meson. Experimentally, the literature only reports, inconclusively,  
the formation of three $\Lambda^+_c$-hypernuclei, observed in a series of emulsion
experiments~\cite{exp1,exp2}. But this situation can change in a few years, with the 
starting of operation of the FAIR facility in Germany and the Hadron Facility at JPARC in Japan. 

There are, however, important differences between NMWD in $\Lambda$-hypernuclei and 
$\Lambda^+_c$-hypernuclei. A first difference comes from the mean lifetimes of the two 
hyperons: $\tau_{\Lambda^+_c} \sim 10^{-3} \, \tau_{\Lambda} $. While the mean lifetime of 
the NMWD~\rf{1.1} is of the same order of magnitude of the full mean lifetime of
$\Lambda$ in free space, no theoretical estimate has been made for the decays \rf{1.2} and 
\rf{1.3}. In addition, while the free-space decay of $\Lambda$ is dominated by the pionic 
channels $\Lambda\go p\pi^-$ and $\Lambda\to n\pi^0$, with other decay channels contributing 
a thousand times less, $\Lambda^+_c$ decays in two semileptonic and numerous hadronic 
channels with ${\sf S}=-1$ final states, having branching ratios of a few percent each.
Also, decays into channels with ${\sf S}=0$ and ${\sf S}=+1$ are Cabibbo-suppressed by factors 
of the order of $10^{-1}{\textendash}10^{-2}$~\cite{PDG}. A second very important difference
concerns the energy liberated in the decays, which is of the order of the mass difference
$\Delta$ of the particles involved in the weak vertex: for the decay~\rf{1.2}, 
$\Delta = M_{\Lambda^+_c}- M_\Lambda=1170.9$~MeV, and for the decay~\rf{1.3}, 
$\Delta = M_{\Lambda^+_c}-\Mass_N=1348.2$~MeV, which should be compared to 
$\Delta = M_{\Lambda}-M_N=177.3$~MeV for the decay~\rf{1.1}. One consequence of such large
energy releases is that nonrelativistic approaches, like those of Refs.~
\cite{Du96,Pa97,It02,It03,It08,Ba02,Kr03,Ba03,Ba05,Ba07,Ba08,Ba10,Kr10,Kr10a,
Go11,Kr14,Kr14a},
become inapplicable for the 
evaluation of NMWD transition matrix elements in charmed hypernuclei. In addition, a large 
energy release also implies that nuclear recoil cannot be neglected in the calculation of
decay rates, particularly for light-weight nuclei. On the other hand, the interactions 
of the fast outgoing nucleons and/or hyperons with the residual nuclear system are expected to play a minor role. 

In the present paper we develop a relativistic formalism for NMWD of hypernuclei within
an independent-particle shell model (IPSM), and discuss the inclusion of recoil. 
Although the use of a relativistic model for the study of the structure of hypernuclei
dates back to the late 1970's with Brockmann and Weise~\cite{BroWei}, so far little is 
known about the impact of a relativistic approach in the evaluation of NMWD rates. The first 
studies started 25 years ago with Ramos~\etal~\cite{Ra91,Ra92}. These authors used single-particle 
bound-state wave functions obtained by solving the Dirac equation with static Lorentz-scalar 
and -vector Woods-Saxon potentials, and transition matrix elements calculated 
with a $(\pi,K)$ OME model. More recently, a similar approach was used by 
Conti~\etal~\cite{Co09,Co09a}, where the nuclear structure was described by a finite-nucleus, 
relativistic, mesonic-mean-field model. An interesting feature of these studies is
that the reported numerical results for the decay rates differ considerably from those obtained
with nonrelativistic approaches {\textemdash} this is true, \emph{e.g.}, for the 
$_\Lambda^{12}$C hypernucleus, as we show in Table~\ref{T3}. Such differences are larger
than one would expect given the moderate energies involved in the decay process. It is also important to
notice that these predictions strongly contradict the experimental data. 

Our aim in the present paper is to set up a relativistic formalism for NMWD with the perspective
of future applications to charmed hypernuclei. In other instances involving nuclear structure calculations 
at low and intermediate energies, it is often more convenient and simpler to use 
a relativistic approach than a nonrelativistic one~\cite{Ri15}; this seems to be also the case for NMWD {\textemdash}  Ref.~\cite{Hagino} presents a very complete review on relativistic approaches for the study of nuclear structure. Although our approach
for the NMWD of hypernuclei shares similarities with the formalism of Refs.~\cite{Ra91,Ra92,Co09,Co09a}, there are noteworthy differences:
\begin{enumerate}
\item Our final expressions for the decay rates do not involve angular momentum 
projection quantum numbers, since they have been summed over in closed form  
using the Racah algebra, which simplifies the numerical calculation;
\item Spectroscopic factors are evaluated in the second quantized formalism, as  done for instance in Ref.~\cite{Na72}, wi\-thout recurring to the technique of coefficients of fractional pa\-rentage (c.f.p.'s), which is the standard antisymmetrization procedure in the first quantization framework, see e.g. Ref.~\cite{de63};
\item We discuss the inclusion of recoil. 
\end{enumerate}
The predictions of our formalism are compared with available data~\cite{Ok05,Ki06,Ki09,Ag14,Sa05} for the NMWD rates of the 
$^{12}_{\,\Lambda}{\rm C}$ hypernucleus. In addition, we
make a detailed comparison with results obtained in nonrelativistic approaches
that include the same ingredients (like short-range correlations and 
OME model); such a comparison between the outcomes of analogous relativistic and nonrelativistic approaches had not been done so far.  

Our formalism is explained in Section~\ref{Sec2} starting from the simplest scenario, corresponding 
to hypernuclei with closed-shell cores and ignoring recoil, in Subsection~\ref{Sec2A}. This part is done in a strictly
relativistic manner, while the next two steps are performed in analogy to nonrelativistic calculations: first, in Subsection~\ref{Sec2B}, we generalize the formulation to hypernuclei with open-shell cores; and 
secondly, in Subsection~\ref{Sec2C}, the recoil effect is discussed. Subsequently, in Section~\ref{Sec3}, our numerical results for the decay rates of $_\Lambda^{12}$C are presented and compared to those of nonrelativistic calculations using a similar model \cite{Ga13,Co14}. They are also compared with those of previous relativistic calculations and confronted  with the 
experimental data, and a few conclusions are drawn. Finally, in Section~\ref{Sec4}, a general summary is given. In Appendices \ref{A}--\ref{C}, some details of the calculation are presented.

\section{Relativistic Decay Rate }
\label{Sec2}
\begin{figure}[h]
\begin{center}
\scalebox{1.5}{
\leavevmode
\epsfxsize = 4cm
\epsfysize = 6cm
\epsffile{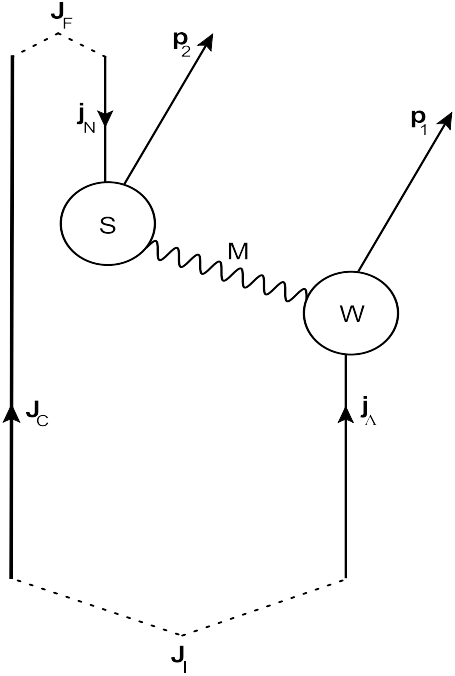}
}
\end{center}
\caption{Diagrammatic representation of the
hypernuclear nonmesonic weak decay from the initial state $\ket{J_I}\equiv\ket{(J_Cj_\Lambda)J_I}$
to the final state $\ket{J_F}\equiv\ket{(J_Cj_N^{-1})J_F}$ while two nucleons with momenta
$\pb_1$ and $\pb_2$ are emitted into the continuum. 
S and W are the strong
and the weak vertices, respectively, and M is a nonstrange meson.
For a strange meson, the natures of the two vertices should be interchanged. 
\label{fig1}}
\end{figure}

To derive the NMWD rate we start from the Fermi Golden Rule.
For a hypernucleus in its ground state with spin $J_I$ and total rest energy $E_{I}$ decaying 
into (i) two free nucleons, with asymptotic kinetic energies ($T_{1}, T_{2}$), spin projections 
(${s_1},{s_2}$), and isospin projections (${t_1},{t_2}$) and (ii) the residual $(A-2)$-system, 
with spin $J_F$, total rest energy $E_{F}$, and kinetic energy of recoil $T_R$, reads
\begin{eqnarray}
\Gamma_{nm} &=&\frac{2\pi }{(2J_I+1)}
\underset{{s_1}{s_2}{t_1}{t_2}}{\underset{M_IJ_FM_F}{\sum}}
\int\frac{d{\bf p}_1}{(2\pi)^3}\frac{d{\bf p}_2}{(2\pi)^3}
\delta(E_I-E_F-\mathcal{E})\nn\\[0.17cm]&\x& |\overline{\M}(\pb_1\pb_2{s_1}{s_2}{t_1}{t_2}
J_FM_F,J_IM_I)|^2,
\label{2.1}
\end{eqnarray}
where $\mathcal{E}=2\Mass_{N}-T_R-T_2-T_1$, $\overline{\M}=(1-P_{12})\M/\sqrt{2}$ is the antisymmetrized and normalized relativistic matrix element that is specified below, $\Mass_N$ is the nucleon mass, and $p_i=\sqrt{E_i^2-\Mass_N^2}$ and $E_i=T_i+\Mass_N$  are the asymptotic momenta and total energies of the outgoing particles ($i=1,2$).
We use unitary, as opposed to co\-variant, normalization for the momentum eigenspinors;
for details see Section 2.2 of Ref.~\cite{Se86}.
We average over the spin projections $M_I$ of the initial hypernucleus and sum over the final spin projections $M_F$.
For the nuclear structure framework, the IPSM is used,
while the dynamics is  des\-cribed by an OME potential  containing
always one weak vertex W and and one strong vertex S, as illustrated in 
Figure~\ref{fig1}.
In the IPSM it is assumed that: (i)
the initial hypernuclear state can be approximated as a $\Lambda$-hyperon in the
single-particle state $j_\Lambda=1s_{1/2}$ weakly coupled to an
$(A-1)$ nuclear core of spin $J_C$ and total rest ener\-gy $E_C$, \ie
$\ket{J_I}\equiv\ket{(J_Cj_\Lambda)J_I}$, having energy 
$E_I= E_C+\varepsilon_{j_\Lambda}+\Mass_\Lambda$; (ii)
the nucleon $N$ inducing the decay is in the single-particle state
$j_N$ ($j\equiv nlj$); (iii) the final residual nuclear states have the form
$\ket{J_F}\equiv\ket{(J_Cj_N^{-1})J_F}$ with energy $E_F= E_C-\varepsilon_{j_N}-\Mass_N$; (iv) the liberated kinetic energy is
\be T_R+T_2+T_1= E_I-E_F-2\Mass_N\equiv \Delta_{j_N} =\Delta  +
\varepsilon_{j_\Lambda} + \varepsilon_{j_N}
\label{2.2}
\ee
where $\Delta=\Mass_\Lambda-\Mass_N=177$ MeV, and the $\varepsilon$'s are single-particle energies.

\subsection{Hypernuclei with doubly-closed shell cores and without recoil}
\label{Sec2A}
Taking the simplest possible case in Eq.~\rf{2.1}, we will start  with
hypernuclei whose cores contain only doubly-closed subshells,
as, for instance, $_\Lambda^{5}$He, $_\Lambda^{13}$C,
$_\Lambda^{17}$O, and we will omit the recoil effect.
Thus, $J_C=0$, $J_I=j_\Lambda$, $M_I=m_\Lambda$, $J_F=j_N$, $M_F=m_N$, 
and the transition amplitude $\M$  is just the two-body T-matrix for the direct OME process. When a pseudoscalar coupling is considered for the strong vertex, one has for the pion plus kaon meson exchange: $\M=\M^\pi+\M^K$, see ~\cite[Eq. (3)]{Ra92} and \cite[Eqs.(7),(45)]{Co09}, with
\br
\lefteqn{
\M^\pi(\pb_1\pb_2{s_1}{s_2}{t_1}{t_2}j_N m_Nj_\Lambda m_\Lambda) \;=\;
}
\nn\\&& 
\int d\xb\, d\yb\, \bar{\psi}_{\pb_1s_1}(\xb)
[\A^\pi(t_1,t_2) - \B^\pi(t_1,t_2) \gamma_5] 
\nn\\&&\times 
{\Psi}_{j_\Lambda m_\Lambda}(\xb)
{\sf \Delta}^\pi(|\xb-\yb|)\bar{\psi}_{\pb_2s_2}(\yb)
\gamma_5{\Psi}_{j_Nm_N}(\yb),
\label{2.4}
\er
and
\br
\lefteqn{
\M^K(\pb_1\pb_2{s_1}{s_2}{t_1}{t_2}j_N m_Nj_\Lambda m_\Lambda) \;=\;
}
\nn\\&& 
\int d\xb\,d\yb\,\bar{\psi}_{\pb_1s_1}(\xb)\gamma_5 
{\Psi}_{j_\Lambda m_\Lambda}(\xb)
{\sf \Delta}^K(|\xb-\yb|)
\nn\\&&\times
\bar{\psi}_{\pb_2s_2}(\yb)
[\A^K(t_1,t_2) -\B^K(t_1,t_2)\gamma_5] {\Psi}_{j_Nm_N}(\yb),
\label{2.5}
\er
where we are using the following definitions
\br
\A^{\pi}(t_1,t_2) &=& G_Fm_\pi^2g_{\pi NN}\tilde{A}^{\pi}(t_{1},t_{2}),
\nn\\
\A^{K}(t_1,t_2) &=& G_Fm_\pi^2 g_{K\Lambda N}\tilde{A}^{K}(t_{1},t_{2}),
\nn\\
\B^{\pi}(t_1,t_2) &=& G_Fm_\pi^2 g_{\pi NN}\tilde{B}^{\pi}(t_{1},t_{2}),
\nn\\
\B^{K }(t_1,t_2) &=& G_Fm_\pi^2 g_{K\Lambda N}\tilde{B}^{K}(t_{1},t_{2}),
\label{2.6}
\er
with $\tilde{A}^{\pi}(t_{1},t_{2})=AI$, 
$\tilde{A}^{K}(t_{1},t_{2})=(IA_1+KA_0)$,
$\tilde{B}^{\pi}(t_{1},t_{2})=BI$, 
$\tilde{B}^{K}(t_{1},t_{2})=(IB_1+KB_0)$
and where $I=\bra{t_1}\mbt_1\ket{t_\Lambda=-\fot}\cdot \bra{t_2}\mbt_2\ket{t_N}$ and
$K=\bra{t_1}1_1\ket{t_\Lambda=-\fot}\, \bra{t_2}1_2\ket{t_N}$ are, respectively,
the isovector and isoscalar isospin factors.
Here, $G_Fm_\pi^2=2.21\x10^{-7}$, with $G_F$ being  the Fermi weak constant and $m_\pi$ the pion mass, while $g_{\pi NN}=13.3$ and $g_{K\Lambda N}=-14.1$  are the strong vertex couplings \cite{Ma89}.
The pion  parity-violating (PV) and parity-conserving (PC)  weak coupling
constants are adjusted to the free $\Lambda$-decay giving, respectively, $A = 1.05$ and $B = -7.15$, while the kaon weak couplings
\br
A_0&=&\frac{C_K^{PV}}{2}+D_K^{PV},\hspace{.1cm}A_1=\frac{C_K^{PV}}{2},
\nn\\
B_0&=&\frac{C_K^{PC}}{2}+D_K^{PC},\hspace{.1cm}B_1=\frac{C_K^{PC}}{2},
\label{2.7}
\er
with $C_K^{PV}=0.76$, $C_K^{PC}=-18.9$, $D_K^{PV}=2.09$, and $D_K^{PC}=6.63$,
have been estimated theoretically \cite{Du96}.

The propagator  $\sDelta^M(|\xb-\yb|)\; (M=\pi,K)$ reads \cite{Co09,Co09a}:
\begin{enumerate}
\item 
\label{item1}
For $({q}_0)^2<m_M^2$,  
\br
\sDelta^M(|\xb-\yb|)&=&
-\frac{\exp(-\sqrt{m_M^2-q_0^2}\,|\xb-\yb|)}{4\pi|\xb-\yb|}
\nn\\&+&\sDelta_F^M(\xb-\yb),
\label{2.8}
\er
where $q_0$ is the energy carried by the exchanged meson, and
\br
\lefteqn{
\sDelta_F^M(|\xb-\yb|) \;=\; 
-\frac{\exp(-\sqrt{\Lambda_M^2-q_0^2}\,|\xb-\yb|)}{4\pi|\xb-\yb|}
}
\nn\\
&+&\frac{\Lambda_M^2-m_M^2}{8\pi\sqrt{\Lambda_M^2-q_0^2}}\exp(-\sqrt{\Lambda_M^2-q_0^2}\,|\xb-\yb|),
\label{2.9}
\er
is the finite-size correction when dipole form factors with cut-off parameters $\Lambda_M$ ($M=\pi,K$) are attached  at each vertex;
\item 
For $m_M^2<({q}_0)^2<\Lambda_M^2$,
\br
\sDelta^M(|\xb-\yb|) &=&
-\frac{\exp(i\sqrt{q_0^2-m_M^2}\,|\xb-\yb|)}{4\pi|\xb-\yb|}\,.
\label{2.10}
\er
Therefore the propagator is now complex \cite{Co09,Co09a} and can have oscillatory behavior, with real and imaginary parts given by
\begin{eqnarray}
\Re\sDelta^M(|\xb-\yb|)&=& 
-\frac{\cos(\sqrt{q_0^2-m_M^2}|\xb-\yb|)}{4\pi|\xb-\yb|},
\label{2.11a}
\\
\Im\sDelta^M(|\xb-\yb|)&=& 
-\frac{\sin(\sqrt{q_0^2-m_M^2}|\xb-\yb|)}{4\pi|\xb-\yb|},
\label{2.11b}
\end{eqnarray}
in this region of transferred energy.
\end{enumerate}

The state of each ejected nucleon, with asymptotic momentum $\pb$ and spin projection $s$, will be approximated by a Dirac plane wave, which is expanded in spherical partial-waves as follows~\cite[Appendix D]{Do85}:
\begin{eqnarray}
\psi_{\pb s}(\rb)&=&\sum_{\kappa m}
\ov{\ph s}{\kappa m}^\ast
\psi_{p\kappa m}(\rb),
\label{2.12a}
\\
\ov{\ph s}{\kappa m}^\ast 
&=&4\pi i^l\sum_{\mu}(l\mu\fot s|jm)Y^*_{l\mu}(\ph),
\label{2.12b}
\\
\psi_{p\kappa m}(\rb)&=&
\bin{f_{p\kappa}(r)\Phi_{\kappa m}(\rh)}
{-ig_{p\kappa}(r)\Phi_{-\kappa m}(\rh)} 
\nn\\
&\equiv&\bin{\psip_{p\kappa m}(\rb)}{-i\psin_{p\kappa m}(\rb)},
\label{2.12c}
\end{eqnarray}
where the radial partial-waves are, in unitary norma\-lization,
\begin{eqnarray}
f_{p\kappa}(r)&=&\sqrt{\frac{E+\Mass_N}{2E}}j_{l_\kappa}(pr)
\label{2.13a}
\\
g_{p\kappa}(r)&=&
-\sgn(\kappa)\sqrt{\frac{E-\Mass_N}{2E}}j_{{\bar l}_\kappa}(pr),
\label{2.13b}
\end{eqnarray}
with $\kappa=\pm1,\pm2,\dots$, $j_\kappa=|\kappa|-1/2$, 
\be 
l_{\kappa} = \left\{ \begin{array}{ll}\kappa & \mbox{for $\kappa>0$} \\
-\kappa-1 & \mbox{for $\kappa<0$} \end{array} \right.,
\label{2.14}
\ee
and ${\bar l}_\kappa=l_{-\kappa}$. 
(To change to covariant normalization, used in Refs.\cite{Ra92,Co09,Co09a}, make the replacement $\sqrt{2E}\go\sqrt{2\Mass_N}$ in Eqs. \rf{2.13a} and \rf{2.13b} and 
insert the factor $\Mass^2_N/(E_1E_2)$ in Eq.\rf{2.1}.)
The angular part is written, in standard notation, as
\br
\Phi_{\kappa m}(\rh)&=&\sum_{s\mu}(l\mu\fot s|jm)Y_{l\mu}(\rh)\chi_s,
\er
and the expansion coefficients 
$\ov{\ph s}{\kappa m}^\ast $ 
fulfill the following relations
\be
\sum_s\int d\ph\, 
\ov{\ph s}{\kappa m}^\ast \ov{\ph s}{\kappa' m'} 
=(4\pi)^2\delta_{\kappa\kappa'}\delta_{mm'},
\label{2.16a}
\ee
\br
\lefteqn{
2\hat{j}^{2}\delta_{jj'}\hspace{-.1cm}\sum_{s m}\int d\ph\, 
\ov{\ph s}{\kappa m}^\ast \ov{\ph s}{\kappa' m} \cdots \;=\;
}
\hspace{2cm}
\nn\\
&&(4\pi)^2\delta_{\kappa\kappa'}\int_{-1}^1 d\cos\theta \cdots,\quad\quad
\label{2.16b}
\er
where we are using the notation $\hat{j}=\sqrt{2j+1}$.
The first of these relations can be easily verified, while the second one is shown
in Appendix~\ref{A}.
The bound-state, single-particle, wave functions read
\br
\Psi_{\kappa m}(\rb)
&=&\frac{1}{r}\bin{F_{\kappa}(r)\Phi_{\kappa m}(\rh)}{-iG_{\kappa}(r)(\rh)\Phi_{-\kappa m}(\rh)}
\nn\\
&\equiv&\bin{\Psip_{\kappa m}(\rb)}{-i\Psin_{\kappa m}(\rb)}.
\label{2.17}
\er
As explained in Appendix~\ref{B}, they are evaluated as in Ref.~\cite[Eq.(16)]{Ho81}.

To simplify the presentation of formulas in the analytical development of Eq.~\rf{2.1}, the intermediate steps
will be exhibited only for $\M^\pi$, which is rewritten as
\br
&&\hspace{-1.1cm}\M^{\pi}(\pb_1\pb_2{s_1}{s_2}{t_1}{t_2}j_N m_Nj_\Lambda m_\Lambda)
\nn\\
&&=\underset{\kappa_2m_2}{\underset{\kappa_1m_1}{\sum}}
\ov{\ph_1 s_1}{\kappa_1 m_1} \ov{\ph_2 s_2}{\kappa_2 m_2}
\nn\\&&\x 
\bra{p_1\kappa_1 m_1t_1p_2\kappa_2m_2t_2}{\sf \Delta}^\pi\ket{\kappa_\Lambda m_\Lambda\kappa_N m_N},
\label{2.18}
\er
where we are using the following compact notation
\br
\lefteqn{
\bra{p_1\kappa_1 m_1t_1p_2\kappa_2m_2t_2}
{\sf \Delta}^\pi\ket{\kappa_\Lambda m_\Lambda\kappa_N m_N}
}
\nn\\
&\equiv&\int d\xb d\yb\, \bar{\psi}_{p_1\kappa_1 m_1}(\xb) 
[\A^\pi(t_1,t_2) - \B^\pi(t_1,t_2)\gamma_5]
\nn\\
&\times&
{\Psi}_{j_\Lambda m_\Lambda}(\xb)
{\sf \Delta}^\pi(|\xb-\yb|)\bar{\psi}_{p_2\kappa_2 m_2}(\yb)\gamma_5{\Psi}_{j_Nm_N}(\yb).\quad{}
\label{2.19}
\er
Introducing these expansions in Eq. \rf{2.1} gives rise to auxiliary quantities such as
\br
S^\pi(p_1t_1,p_2t_2)&\equiv &
\underset{s_1s_2}{\underset{m_\Lambda m_N}{\sum}}
\int d\ph_1 d\ph_2\delta(\Delta_{j_N}-T_1-T_2-T_R)
\nn\\&\times& |\M^\pi(\pb_1\pb_2{s_1}{s_2}{t_1}{t_2}j_N m_Nj_\Lambda m_\Lambda)|^2,
\label{2.20}
\er
in which we evaluate all the  summations over angular momentum projection quantum numbers and angular integrations.
Neglecting recoil, \ie setting $T_R=0$, 
we can use
\rf{2.16a} for both outgoing particles getting
\br
&&\hspace{-.8cm}S^\pi(p_1t_1,p_2t_2)
=(4\pi)^4\sum_{m_\Lambda m_N}
\underset{\kappa_2m_2}{\underset{\kappa_1m_1}{\sum}}
\delta(\Delta_{j_N}-T_1-T_2)
\nn\\&&\hspace{.1cm}\times
\left| \bra{p_1\kappa_1 m_1t_1p_2\kappa_2m_2t_2}{\sf \Delta}^\pi\ket{\kappa_\Lambda m_\Lambda\kappa_N m_N}\right|^2.
\label{2.21}
\er
Now we perform the angular momentum couplings $\vec{J}=\vec{j}_\Lambda+\vec{j}_N$ and $\vec{J}'=\vec{j}_1+\vec{j}_2$.  
As ${\sf \Delta}^\pi$ is rotationally invariant, it turns out that $J = J'$, which leads to
\br
\lefteqn{
\bra{p_1\kappa_1 m_1t_1p_2\kappa_2m_2t_2}{\sf \Delta}^\pi\ket{\kappa_\Lambda m_\Lambda\kappa_N m_N} 
}
\nn\\
&=&\sum_{JM}
\bra{p_1\kappa_1t_1 p_2\kappa_2t_2J}{\sf \Delta}^\pi
\ket{\kappa_\Lambda \kappa_NJ}
\nn\\
&&\times
(j_1m_1j_2m_2|JM)\,
(j_\Lambda m_\Lambda j_N m_N|JM),
\label{2.22}
\er
and
\br
S^\pi(p_1t_1,p_2t_2) &=&
(4\pi)^4\sum_{\kappa_1\kappa_2J}\hat{J}^{2}\delta(\Delta_{j_N}-T_1-T_2)
\nn\\&\times&
\left|\bra{p_1\kappa_1t_1 p_2\kappa_2t_2J}{\sf \Delta}^\pi\ket{\kappa_\Lambda \kappa_NJ}\right|^2,
\label{2.23}
\er
where the coupled matrix element of the pion propagator is explicitly
given by
\begin{widetext}
\br
\bra{p_1\kappa_1t_1 p_2\kappa_2t_2J}{\sf \Delta}^\pi\ket{\kappa_\Lambda \kappa_NJ}
&=&
-i\int d\xb\, d\yb 
\Big\{
\big[\;
\A^\pi(t_1,t_2)
\big(\,
{\psip}^*_{p_1\kappa_1}(\xb){{\Psip}_{\kappa_\Lambda }}(\xb) - {\psin}^*_{p_1\kappa_1}(\xb){{\Psin}^*_{\kappa_\Lambda}}(\xb)
\,\big)
\nn\\
&+& 
\B^\pi(t_1,t_2)
\big(\,
{\psip}^*_{p_1\kappa_1 }(\xb){{\Psin}_{\kappa_\Lambda}}(\xb) + {\psin}^*_{p_1\kappa_1}(\xb){{\Psip}_{\kappa_\Lambda}}(\xb)
\,\big)
\;\big]\;
{\sf \Delta}^\pi(|\xb-\yb|) 
\nn\\
&\x&
\big(\,
{\psip}^*_{p_2\kappa_2}(\yb) {{\Psin}_{\kappa_N}}(\yb) + 
{\psin}^*_{p_2\kappa_2 }(\yb){{\Psip}_{\kappa_N }}(\yb)
\,\big)
\Big\}_{(j_1j_2;j_\Lambda j_N;J)} \,,
\label{2.24}
\er
\end{widetext}
with the above mentioned angular momentum couplings indicated in the last index.

At this point it is convenient to perform  the tensor expansion of the propagators ${\sf\Delta}^\pi(|\xb-\yb|)$  in the way done  by de-Shalit and Talmi ~\cite[Sec. 21]{de63} for two-body interactions, \ie
\br
{\sf \Delta}^\pi(|\xb-\yb|)=\sum_{L}{\sf \Delta}^\pi_L(x,y)[Y_L(\xh)_L\cdot Y(\yh)_L]
\label{2.25}
\er
where
\be
{\sf \Delta}^\pi_{L}(x,y)=2\pi \int {\sf \Delta}^\pi(|\xb-\yb|)P_L(\cos\theta_{xy})d(\cos\theta_{xy})
\label{2.26}
\ee
and
\br
\lefteqn{
\bra{\kappa_1\kappa_2 J} 
[Y_{L}(\xh)\cdot Y_{L}(\yh)] 
\ket{\kappa_\Lambda\kappa_{ N}J}
\;=\; (-)^{j_2+j_{\Lambda}+J}
}
\nn\\
&\times&
\sixj{j_1}{j_2}{J}{j_{ N}}{j_{\Lambda}}{L}
\Bra{\kappa_1}Y_L\Ket{\kappa_\Lambda}
\Bra{\kappa_2}Y_L\Ket{\kappa_{ N}} \,.
\hspace{1cm}{}
\label{2.27}
\er
It is then easy  to  demonstrate  that
\br
\lefteqn{
\bra{\kappa_1 p_1t_1\kappa_2p_2t_2J}{\sf \Delta}^\pi\ket{\kappa_\Lambda \kappa_NJ}
\;=\;\sum_{L}(-)^{j_2+j_{\Lambda}+J}
}
\nn\\&\times&
\sixj{j_1}{j_2}{J}{j_{ N}}{j_{\Lambda}}{L}
\bra{\kappa_1p_1t_1\kappa_2p_2t_2}
{\sf \Delta}^\pi_{L}\ket{\kappa_\Lambda\kappa_N} \,,
\label{2.28}
\er
where
\br
&&\hspace{-1.2cm}\bra{\kappa_1p_1t_1\kappa_2p_2t_2}
{\sf \Delta}^\pi_{L}\ket{\kappa_\Lambda\kappa_N}
\nn\\&&\hspace{-.8cm}
=\int x y dx dy \, [\B^\pi(t_1,t_2) B^L_{\kappa_1\kappa_\Lambda}(xp_1) 
\nn\\&&\hspace{-.8cm}
{} -i\A^\pi(t_1,t_2) A^L_{\kappa_1\kappa_\Lambda}(xp_1)] \, 
{\sf \Delta}^\pi_{L}(x,y)C^L_{\kappa_2\kappa_N}(yp_2)
\label{2.29}
\er
with
\br
A^L_{\kappa\kappa_\Lambda}(rp) &=&
[f_{p\kappa}(r)F_{\kappa_\Lambda}(r)-g_{p\kappa}(r)G_{\kappa_\Lambda}(r)]
\nn\\&&\times
\Bra{\kappa}Y_{L}\Ket{\kappa_\Lambda},
\nn\\
B^L_{\kappa\kappa_\Lambda}(rp) &=&
[f_{p\kappa}(r)G_{\kappa_\Lambda}(r)+g_{p\kappa}(r)F_{\kappa_\Lambda}(r)]
\nn\\&&\times
\Bra{-\kappa}Y_{L}\Ket{\kappa_\Lambda},
\nn\\
C^L_{\kappa\kappa_N}(rp) &=&
[f_{p\kappa}(r)G_{\kappa_N}(r)+g_{p\kappa}(r)F_{\kappa_N}(r)]
\nn\\&&\times
\Bra{-\kappa}Y_{L}\Ket{\kappa_N}.
\label{2.30}
\er
The reduced matrix elements
\br
\Bra{\kappa}Y_L\Ket{\kappa'}&=&(4\pi)^{-1/2}(-)^{j-1/2}\hat{j}\hat{j}'\hat{L}
\nn\\ &\times&
\threej{j}{L}{j'}{-\fot}{0}{\fot}\frac{1+(-)^{l+l'+L}}{2}
\label{2.31a}
\er
and
\br
\Bra{-\kappa}Y_L\Ket{\kappa'}&=&(4\pi)^{-1/2}(-)^{j-1/2}\hat{j}\hat{j}'\hat{L}
\nn\\ &\times&
\threej{j}{L}{j'}{-\fot}{0}{\fot}\frac{1+(-)^{{\bar l}+l'+L}}{2}
\label{2.31}
\er
fulfill the symmetry relations $\Bra{\kappa}Y_L\Ket{\kappa'}=\Bra{\kappa'}Y_L\Ket{\kappa}$, and
$\Bra{\kappa}Y_L\Ket{-\kappa'}=\Bra{-\kappa}Y_L\Ket{\kappa'}$.

The $K$ meson is incorporated through
the substitution ${\sf \Delta}^\pi_L\go {\sf \Delta}_L={\sf \Delta}^\pi_L+{\sf \Delta}^K_L$ in \rf{2.29}, with
\br
&&\hspace{-.8cm}\bra{\kappa_1p_1t_1\kappa_2p_2t_2}
{\sf \Delta}^K_{L}\ket{\kappa_\Lambda\kappa_N}
\nn\\&&\hspace{-.6cm}
=\int x y dx dy \,
B^L_{\kappa_1\kappa_\Lambda}(xp_1) 
{\sf\Delta}^K_{L}(x,y) \,
\nn\\ &&\hspace{-.6cm}
\times
[\B^K(t_1,t_2) C^L_{\kappa_2\kappa_N}(yp_2)
-i\A^K(t_1,t_2)
D^L_{\kappa_2\kappa_N}(yp_2)] ,
\label{2.32}
\er
where
\br
D^L_{\kappa\kappa_N}(rp) &=&
[f_{p\kappa}(r)F_{\kappa_N}(r)-g_{p\kappa}(r)G_{\kappa_N}(r)]
\nn\\&&\times
\Bra{\kappa}Y_{L}\Ket{\kappa_N}.
\label{2.30a}
\er
Clearly the above substitution must be accompanied by the replacement ${\sf \Delta}^\pi\go {\sf \Delta}={\sf \Delta}^\pi+{\sf \Delta}^K$ in \rf{2.28}, giving
\br
\lefteqn{
\bra{\kappa_1p_1t_1 \kappa_2p_2t_2J}{\sf \Delta}\ket{\kappa_\Lambda \kappa_NJ}
\;=\; \sum_{L}(-)^{j_2+j_{\Lambda}+J}
}
\nn\\ &\times&
\sixj{j_1}{j_2}{J}{j_{ N}}{j_{\Lambda}}{L}
\bra{\kappa_1p_1t_1\kappa_2p_2t_1}{\sf \Delta}_{L}\ket{\kappa_\Lambda\kappa_N}.
\hspace{1cm}{}
\label{2.33}
\er
Finally, from \rf{2.1},
\begin{eqnarray}
\Gamma_N &=&\frac{8}{\pi }
\underset{\kappa_1\kappa_2J}{\underset{j_Nt_1t_2}{\sum}}\,
\frac{\hat{J}^{2}}{\hat{j}_\Lambda^{2}}\int\, p_1^2dp_1p^2_2dp_2\,
\delta(\Delta_{j_N}-T_1-T_2)
\hspace{.5cm}
\nn\\ &\times&
\left|\bra{\overline{\kappa_1p_1t_1\kappa_2p_2t_2J}}
{\sf \Delta}\ket{\kappa_\Lambda\kappa_NJ}\right|^2,
\label{2.34}
\end{eqnarray}
where
\br
\ket{\overline{\kappa_1p_1t_1\kappa_2p_2t_2J}}&=&\frac{1}{\sqrt{2}}
\big(\,\ket{\kappa_1p_1t_1\kappa_2p_2t_2J}
\nn\\ &-&
(-)^{j_1+j_2-J}\ket{\kappa_2p_2t_2\kappa_1p_1t_1J}\,\big)
\label{2.35}
\er
stand for the antisymmetrized and normalized two-particle wave functions  with the isospins included.
The isospin factors for the direct  and exchange terms of the matrix-element in Eq. \rf{2.34} are listed in Table \ref{T1}. 
\begin{table}[h]
\centering \caption{Isospin factors the for direct (D) and exchange (E) terms of the matrix-element in Eq. \rf{2.34}.
} \label{T1}
\bigskip
\small\addtolength{\tabcolsep}{-3pt}
\begin{tabular}{ccccc}
\hline \noalign{\smallskip}
&$\hspace{2cm}I$ &$\hspace{2cm}$ &$\hspace{2cm}K$ &$\hspace{2cm}$ \\
\hline
            &$n$ &$p$            &$n$ &$p$ \\
\noalign{\smallskip}
\hline
\noalign{\smallskip}
D&$1$&$-1$&$1$&$1$\\
E&$1$&$2$&$1$&$0$\\
\noalign{\smallskip}
\hline
\end{tabular}
\end{table}

It is worth noting that  the matrix elements
$\bra{\overline{\kappa_1p_1t_1\kappa_2p_2t_2J}}
{\sf \Delta}\ket{\kappa_\Lambda\kappa_NJ}$ 
are in general complex, as seen from Eqs.~\rf{2.29} and \rf{2.32}.
However, in the usual regime of item~\ref{item1} on page~\pageref{item1}, 
they are always either real or purely imaginary because 
there is no set of quantum numbers for which parity-conserving and parity-violating contributions interfere with each other.

To exploit  the delta function
in \rf{2.34} we make use of the relation
\br
p_i^2 dp_i&=&E_i\sqrt{E_i^2-\Mass_N^2}dE_i
\nn\\ &=&(\Mass_N+T_i)\sqrt{T_i(2\Mass_N+T_i)}dT_i
\label{2.36}
\er
and get
\begin{eqnarray}
\hspace{-.5cm}
\Gamma_N &=&\frac{8}{\pi }
\underset{\kappa_1\kappa_2J}{\underset{j_Nt_1t_2}{\sum}}\,
\frac{\hat{J}^{2}}{\hat{j}_\Lambda^{2}}\int dT_1dT_2\,\,
\delta(\Delta_{j_N}-T_1-T_2)
\nn\\ &\times &
\rho(T_1,T_2)\left|\bra{\overline{\kappa_1p_1t_1\kappa_2p_2t_2J}}
{\sf \Delta}\ket{\kappa_\Lambda\kappa_NJ}\right|^2 ,
\label{2.37}
\end{eqnarray}
where
\br
\rho(T_1,T_2)&=&(\Mass_N+T_1)\sqrt{T_1(2\Mass_N+T_1)}
\nn\\ &\times &
(\Mass_N+T_2)\sqrt{T_2(2\Mass_N+T_2)}.
\label{2.38}
\er
After integrating over $T_2$ we are  left with the $T_1$ integration only,
\begin{eqnarray}
\hspace{-.5cm}
\Gamma_N &=&\frac{8}{\pi }\underset{\kappa_1\kappa_2J}{\underset{j_Nt_1t_2}{\sum}}
\,\frac{\hat{J}^{2}}{\hat{j}_\Lambda^{2}}\int_0^{\Delta_{j_N}}dT_1
\rho(T_1,T_2)
\nn\\ &\times &
\left|\bra{\overline{\kappa_1p_1t_1\kappa_2p_2t_2J}} {\sf \Delta}\ket{\kappa_\Lambda\kappa_NJ}\right|^2{\Big{|}}_{T_2=\Delta_{j_N}-T_1},
\label{2.39}
\end{eqnarray}
where
\begin{eqnarray}
&&\hspace{-1.2cm}\bra{\overline{\kappa_1p_1t_1\kappa_2p_2t_2J}} {\sf \Delta}\ket{\kappa_\Lambda\kappa_NJ}
\nn\\ &&\hspace{-.6cm}=\frac{1}{\sqrt{2}}
(\bra{\kappa_1p_1t_1 \kappa_2p_2t_2J}{\sf \Delta}\ket{\kappa_\Lambda \kappa_NJ}
\nn\\
&&\hspace{-.6cm}-(-)^{j_1+j_2-J}\bra{\kappa_2p_2t_2 \kappa_1p_1t_1J}{\sf \Delta}\ket{\kappa_\Lambda \kappa_NJ}).
\label{2.40}
\end{eqnarray}
The direct matrix element is given by \rf{2.33} and the exchange one is obtained through the transposition  $(\kappa_1,p_1,t_1)\lra(\kappa_2,p_2,t_2)$.

\subsection{Hypernuclei  with open-shell cores and without recoil}
\label{Sec2B}

So far everything was done in the strict framework of relativistic physics.
In what follows we will make use of analogies with nonrelativistic calculations.
From previous  works ~\cite{Ba08,Ba10,Kr10,Kr10a,Go11,Kr14,Kr14a} done by our group, we know that to describe the hypernuclei with open-shell cores within the IPSM it is enough to do  the following replacement in Eq.\rf{2.39}
\br
\frac{\hat{J}^{2}}{\hat{j}_\Lambda^{2}}\go
F^{j_N}_J
\label{2.41a}
\er
where the spectroscopic factor is given by
\br
\hspace{-.5cm}F^{j_N}_J&=&\hat{J}^{-2}\sum_{J_F} |\Bra{J_I}\left( a_{j_N}^\dag a_{j_\Lambda }^\dag\right)_{J}\Ket{J_F}|^2
\nn\\
&=&\hat{J}^{2}\sum_{J_F}\sixj{J_C}{J_I}{j_\Lambda}{J}{j_N}{J_F}^2|\Bra{J_C}a_{j_N}^\dag\Ket{J_F}|^2.
\label{2.41}
\er
As previously mentioned, to evaluate the spectroscopic amplitudes $\Bra{J_C}a_{j_N}^\dag\Ket{J_F}$, instead of employing the c.f.p.'s \cite{de63} that have been thoroughly used in, both nonrelativistic~\cite{Pa97}, and relativistic~\cite{Ra92,Co09,Co09a} calculations, we use the second quantization formalism.
In  \rf{2.41} the summation  goes only over the values of $J_F$ that fulfill the constraint $|J_C-j_N|\le J_F\le J_C+j_N$.
The values for $J_I$ and $J_C$ are  taken from experimental data and, for most hypernuclei of interest, are listed in Table I of Ref.~\cite{Kr10}. The resulting factors $F^{j_N}_J$ are listed in Table II of the same paper.

Therefore, when the recoil effect is not taken into account,
the NMWD transition rate in open shell hypernuclei reads
\begin{eqnarray}
\Gamma_N &=&\frac{8}{\pi }
\underset{\kappa_1\kappa_2J}{\underset{j_Nt_1t_2}{\sum}}
F^{j_N}_J
\int_0^{\Delta_{j_N}}dT_1 \rho(T_1,T_2)
\nn\\&\times&
\left|\bra{\overline{\kappa_1p_1t_1\kappa_2p_2t_2J}} 
{\sf \Delta} \ket{\kappa_\Lambda\kappa_NJ}\right|^2
{\Big{|}}_{T_2=\Delta_{j_N}-T_1}.
\label{2.42}
\end{eqnarray}
We note that, while Eq.~\rf{2.39} is only valid for doubly-closed-shell hypernuclei, 
Eq.~\rf{2.42} is  valid for both  closed- and open-shell hypernuclei.

\subsection{Inclusion of the recoil effect }
\label{Sec2C}

As seen above, when the recoil is neglected one can perform first the full angular integration
$\int d\ph_1\int d\ph_2$, leading to a great simplification of the resulting expression.
It is self-evident that this cannot be done anymore in the
presence of the recoil energy
\br
E_R&=&\sqrt{\Mass_R^2+p_1^2+p_2^2+2p_1p_2\cos\theta_{12}},
\label{2.43}
\er
where $\Mass_R$ is the relativistic rest mass of the recoiling nucleus. However,
once the hypernucleus is unpolarized (and unaligned), there is no preferred
axis along which to orient vectors. 
Therefore, we can choose to orient $\pb_2$ with respect to $\pb_1$ and write
\br
\lefteqn{
\int d\ph_1\int d\ph_2 \cdots \;=\;
\int d\ph_1\int d\ph_{12} \cdots 
}
\nn\\
&=& 
\int d\phi_1 \int d\cos\theta_1  
\int d\phi_{12} \int d\cos\theta_{12} \cdots . 
\label{change}
\er
Consequently, we can use \rf{2.16a} for integration on $\ph_1$
and \rf{2.16b} for integration on $\ph_{12}$, with the result that, as shown in 
Appendix \ref{C}, instead of \rf{2.23} we have now
\br
\lefteqn{
S^\pi(p_1t_1,p_2t_2)
}
\nn\\
&=& \frac{(4\pi)^2}{2}\sum_{\kappa_1\kappa_2J}\hat{J}^2
\int d\cos\theta_{12}\, \delta(\Delta_{j_N}-T_1-T_2-T_R)
\nn\\
&\times&
\left|\bra{p_1\kappa_1t_1 p_2\kappa_2t_2J}{\sf \Delta}^\pi\ket{\kappa_\Lambda \kappa_NJ}\right|^2.
\label{2.44}
\er
From comparison with \rf{2.23} one concludes  that the  results developed so far  hold valid even when the recoil effect is included, as long as one makes the replacement:
\br
\lefteqn{
\int p_1^2dp_1p^2_2dp_2\,
\delta(\Delta_{j_N}-T_1-T_2) \cdots
}
\nn\\
&\go&\fot\int d\cos\theta_{12}\, p_1^2dp_1p^2_2dp_2
\nn\\&\times&
\delta(\Delta_{j_N}-T_1-T_2-T_R) \cdots\,.
\label{2.45}
\er

For the sake of convenience we
will work here with  the nonrelativistic  limit for the kinetic energy of recoil, \ie with
\br
\hspace{-.5cm}T_R=E_R-\Mass_R&\cong&\frac{p_1^2+2p_2^2+p_1p_2\cos\theta_{12}}{2\Mass_R}
\nn\\
&\cong&\frac{\Mass_N}{\Mass_R}({T_1+T_2-2\sqrt{T_1T_2}\cos\theta_{12}})
\label{2.46}
\er
which  we consider to be good enough for the present purposes. Moreover, we neglect the binding energy
of the recoiling nucleus, and take $\Mass_R=\Mass_N(A-2)$.
The transition rate becomes then
\begin{eqnarray}
\Gamma_N &=& \frac{4}{\pi }
\underset{\kappa_1\kappa_2J}{\underset{j_Nt_1t_2}{\sum}}
F^{j_N}_J
\int dT_1\, dT_2\, d\cos\theta_{12}\, \rho(T_1,T_2)
\nn\\
&\x&
\left|\bra{\overline{\kappa_1p_1t_1\kappa_2p_2t_2J}} 
{\sf \Delta}\ket{\kappa_\Lambda\kappa_NJ}\right|^2
\nn\\
&\times&
\delta(\Delta_{j_N}-T_R-T_1-T_2).
\hspace{2cm}{}
\label{2.47}
\end{eqnarray}
To perform the integration on $T_2$ we introduce an au\-xiliary variable $x$, defined as  $T_2=p^2_2/(2\Mass_N)\equiv x^2$, \ie
\br
\lefteqn{
\delta(T_2+T_1+T_R-\Delta_{j_N})dT_2 \;=\; \frac{A-2}{A-1}
}
\nn\\
&\times&
\delta\big( x^2+T_1 -\Delta_{j_N}\frac{A-2}{A-1} 
-\frac{2x\cos\theta_{12}}{A-1}\, \sqrt{T_1} \big)\, 2xdx
\nn\\
&=&\frac{A-2}{A-1} 
\frac{2xdx}{|x^+-x^-|}\left[\delta(x-x^+)+\delta(x-x^-)\right],
\hspace{1cm}{}
\label{2.48}
\er
where
\br
x^\pm&=&\frac{\sqrt{T_1}\cos\theta_{12}}{A-1}
\nn\\
&\pm&
\sqrt{\frac{T_1\cos^2\theta_{12}}{(A-1)^2}+\Delta_{j_N}\frac{A-2}{A-1}-T_1}.
\label{2.49}
\er
Therefore
\begin{eqnarray}
\Gamma_N &=& \frac{8}{\pi }\frac{A-2}{A-1}\underset{\kappa_1\kappa_2J}{\underset{j_Nt_1t_2}{\sum}}F^{j_N}_J
\nn\\&\times&
\int dT_1d\cos\theta_{12}dx  x \rho(T_1,T_2)
\nn\\
&\x&
\left|\bra{\overline{\kappa_1p_1t_1\kappa_2p_2t_2J}} 
{\sf \Delta}\ket{\kappa_\Lambda\kappa_NJ}\right|^2
\nn\\&\times&
\frac{\delta(x-x^+)+\delta(x-x^-)}{|x^+-x^-|}.
\label{2.50}
\end{eqnarray}
After integrating on $x$ one gets
\begin{eqnarray}
\Gamma_N &=&
\frac{4}{\pi }\underset{\kappa_1\kappa_2J}{\underset{j_Nt_1t_2}{\sum}}F^{j_N}_J
\nn\\&\times&
\int dT_1d\cos\theta_{12}
\Big\{\Big[
\rho'(T_1,T_2,\cos\theta_{12},\Delta_{j_N})
\nn\\&\times&
\left|\bra{\overline{\kappa_1p_1t_1\kappa_2p_2t_2J}} 
{\sf \Delta}\ket{\kappa_\Lambda\kappa_NJ}\right|^2\Big]_{x\go x^+}
\nn\\
&+& \Big[ \cdot \Big]_{x\go x^-} \Big\},
\label{2.51}
\end{eqnarray}
where
\br
\lefteqn{
\rho'(T_1,T_2,\cos\theta_{12},\Delta_{j_N}) \;=\;
}
\nn\\&&
\frac{x(A-2)\rho(T_1,T_2)}
{\sqrt{T_1\cos^2\theta_{12}+\Delta_{j_N}(A-2)(A-1)-T_1(A-2)^2}}.
\hspace{.5cm}{}
\label{2.52}
\end{eqnarray}

It might be useful to mention here that:
\bit
\item In the analogous nonrelativistic formulation, it has been shown  numerically that the contribution corresponding to the second term
in Eq.~\rf{2.51} is negligibly small compared to that of the first~\cite{Kr10}.
Whether this also occurs in relativistic calculations must still be verified.

\item In the limit $A\go \infty$, the result \rf{2.42} is recovered. Indeed, once
\brn
x^\pm&\Agoinfty&
\pm\sqrt{\Delta_{j_N}-T_1},
\label{2.53}
\ern
$x^-$ becomes unphysical. Therefore, the only contribution comes from the first term in \rf{2.51}, and, as can be seen from \rf{2.52},
\brn
&&\int d\cos\theta_{12}\rho'(T_1,T_2,\cos\theta_{12},\Delta_{j_N})
\nn\\
&&\hspace{0.6cm}\Agoinfty 2\rho(T_1,T_2).
\label{2.54}
\ern
\eit

\section{Numerical results}
\label{Sec3}

We present here our results for the NMWD rates of $^{12}_\Lambda$C. 
The recoil effect has been neglected since we have learned in our previous nonrelativistic calculations \cite{Ba08,Ba10,Kr10,Go11,Kr14,Kr14a} that, although it is relevant for the energy distribution of emitted particles in very light systems, such as s-shell hypernuclei, and is crucial for the angle distribution in general, it is less important for the integrated rates.
Therefore Eq.~\rf{2.42} has been used.

Two approaches have been tested for the propagators $\sDelta^M(|\xb-\yb|)$,
both based on the fact that the ranges of Yukawa-like baryon-baryon forces within hypernuclei depend not only on the intermediate meson mass but also on the baryon masses, as stated in ~\cite[Appendix G]{Ba01}, namely,
\begin{description}
\item[RA1] This is the standard approach in nonrelativistic calculations~\cite{Ba02,Ba03,Sas05}, where the energy $q_0$ carried by the exchanged meson is constant and always smaller that the meson mass $m_M$, having the value  $q_0=\Delta/2=88.5$ MeV. This implies that  the factor 
$\sqrt{m_M^2-q_0^2}$ in  \rf{2.8} is taking the place of the effective mass $\tilde{m}_M=\sqrt{m_M^2-\Delta^2/4}$.
\item[RA2] This is the approach introduced in Refs.~\cite{Co09,Co09a}, which is more appropriate for relativistic calculations, where $q_0$ is evaluated for each value of the kinetic energy $T_1$, with direct  and exchange energies being respectively $q_0^D=\Delta+\varepsilon_{j_\Lambda}-T_1$, and $q_0^E=T_1-\varepsilon_{j_N}$. 
Once for the NMWD in $\Lambda$-hypernuclei the energy transfer
is of the order $50-150$ MeV, $q_0$  can be larger than $m_\pi$ and the factor $\sqrt{m_\pi^2-q_0^2}$ can become complex. Therefore, in the case of the $\pi$ meson, besides making use of Eqs.~\rf{2.8} and \rf{2.9}, one also needs 
Eq.~\rf{2.10}.
We are particularly interested in this approach, since, as mentioned above, the transfer of energy in the NMWD of charmed nuclei can reach much higher values.
\end{description}

Initial and final short range correlations (SRC) were included as in the nonrelativistic case, \ie by ma\-king the substitution
\be
\Delta^M(r) \go g_f(r)\Delta^M(r)g_i(r)
\label{3.1}\ee
in the tensor expansion \rf{2.25},
where $r\equiv|\xb-\yb|$ and 
\br
g_i(r)&=&\left( 1 -  e^{-r^2 / \alpha^2} \right)^2 + \beta r^2
e^{-r^2 / \gamma^2},
\nn\\
g_f(r)&=&1-j_0(q_c r)
\label{3.2}
\er
are, respectively,  the initial and final SRC,
with $\alpha=0.5$~fm, $\beta=0.25$~fm$^{-2}$, and  $\gamma= 1.28$~fm,
and $q_c=3.93$~fm$^{-1}$ \cite{Pa97,Ba02,Ba03,Ra92,Co09,Co09a}.
The dipole form-factor cutoffs $\Lambda_\pi= 1.3$~GeV and  $\Lambda_K=1.2$~GeV
are also the same as in these works.
\begin{table}[t]
\vspace{-.25cm}
\caption{ Comparison between the nonrelativistic (NR) and relativistic  results for $\Gamma_{n}$, $\Gamma_{p}$, $\Gamma_{nm}=\Gamma_{n}+\Gamma_{p}$, and $\Gamma_{n}/\Gamma_{p}$ in ${}^{12}_{\Lambda}\textrm{C}$, for different OME models, \ie the $\pi$ and  $\pi+K$ exchanges without and with SRC. To allow a more detailed comparison, parity-conserving (PC) and parity-violating (PV) parts of $\Gamma_{n}$ and $\Gamma_{p}$ are given separately. As explained in the text, two approximations were used for the propagators in the relativistic calculations, namely RA1 and RA2. All results are in units of the free $\Lambda$ decay rate $\Gamma_\Lambda=2.50\times 10^{-12}$ MeV.}
\vspace{-.3cm}
\label{T2}
\begin{center}
\begin{tabular*}{\columnwidth}{p{1.3cm}@{\extracolsep{\fill}}|ccccccccc}
\hline \hline\\[-.3cm]
 Model  &
$ \Gamma^{(\text{PC})}_{n}  $&
\hspace{.07cm}  $ \Gamma^{(\text{PV})}_{n}  $\hspace{.07cm}  &
\hspace{.07cm}  $ \Gamma^{(\text{PC})}_{p}  $\hspace{.07cm}  &
\hspace{.07cm}  $ \Gamma^{(\text{PV})}_{p}  $\hspace{.07cm}  &
\hspace{.07cm} $ \Gamma_{n}/\Gamma_{p} $\hspace{.07cm}&
\hspace{.07cm}  $ \Gamma_{nm} $\\
\hline
NR \\
No SRC\\
 ${\pi}$  &
$0.1084$ &
$0.1592$ &
$0.8717$ &
$0.4017$ &
$0.2102$ &
$1.5410$
\\[0.06cm]
${\pi} + {K} $  &
$0.0286$ &
$0.1851$ &
$0.3748$ &
$0.3550$ &
$0.2929$ &
$0.9434$
\\[0.06cm]
SRC\\
${\pi}$  &
$0.0122$ &
$0.1753$ &
$0.8062$ &
$0.4475$ &
$01495$ &
$1.4412$
\\[0.06cm]
\hspace{-.1cm}  ${\pi} + {K} $  &
$0.0161$ &
$0.1945$ &
$0.3822$ &
$0.3985$ &
$0.2697$ &
$0.9913$
\\[0.06cm]
\hline\hline
RA1\\
No SRC\\
${\pi}$  &
$0.1207$ &
$0.1531$ &
$0.6894$ &
$0.3309$ &
$0.2683$ &
$1.2941$
\\[0.06cm]
${\pi} + {K}$ &
$0.0555$ &
$0.2394$ &
$0.3955$ &
$0.3825$ &
$0.3790$ &
$1.0729$
\\[0.06cm]
SRC\\
${\pi}$ &
$0.0969$&
$0.1226$&
$0.5210$&
$0.2447$&
$0.2866$&
$0.9853$
\\[0.06cm]
 ${\pi} + {K}$  &
$0.0563$ &
$0.1648$ &
$0.3503$ &
$0.2691$ &
$0.3569$ &
$0.8405$
\\[0.06cm]
\hline
RA2\\
No SRC\\
$\pi$    &
$0.1692$ &
$0.2199$ &
$0.7875$ &
$0.4550$ &
$0.3131$ &
$1.6316$
\\[0.06cm]
 $\pi + K$&
$0.0988$ &
$0.3104$ &
$0.4771$ &
$0.5080$ &
$0.4154$ &
$1.3943$
\\[0.06cm]
SRC\\
$\pi$    &
$0.1454$ &
$0.1865$ &
$0.6209$ &
$0.3619$ &
$0.3377$ &
$1.3147$
\\[0.06cm]
$\pi + K$&
$0.1002$ &
$0.2317$ &
$0.4361$ &
$0.3874$ &
$0.4030$ &
$1.1554$\\
\hline\hline
\end{tabular*}
\end{center}
\end{table}
We present here two different sorts of comparisons involving our results for the decay rates of $^{12}_\Lambda$C.
First, in Table~\ref{T2}, the two relativistic calculations RA1 and RA2 are compared with each other, and also with the analogous nonrelativistic (NR) calculation using the RA1 approach for the propagator. 
The NR calculation is analogous to the relativistic ones in the sense 
that it uses the same OME model, the same SRC, the same single-particle energies, and single-particle wave functions of a harmonic oscillator potential with size parameter $b=1.60\,\mathrm{fm}$, which gives the same root-mean-square radius for the initial hypernucleus.
We show the decay rates $\Gamma_{n}$ and $\Gamma_{p}$, the total one-nucleon-induced nonmesonic decay rates $\Gamma_{nm}=\Gamma_{n}+\Gamma_{p}$, and the ratios $\Gamma_{n}/\Gamma_{p}$ within different OME models, namely, the $\pi$ and $(\pi,K)$ exchanges without and with SRC. Clearly, the relativistic calculations were evaluated in the laboratory frame of reference (LFR). Therefore, we confront them with NR calculations that also were done in the LFR.
These, in turn, have been shown elsewhere~\cite{Ga13,Co14} to nicely agree with the NR evaluation within center-of-mass frame (CMF).
It is not possible here to se\-parate the decay rates $\Gamma_{n}$ and $\Gamma_{p}$ in the usual Block-Dalitz channels~\cite{Bl63}
\br
\begin{array}[b]{lll}
\hspace{-.6cm}
{\sf a} \doteq ~ ^1\mathrm{S}_0\go ^1\!\mathrm{S}_0,
\quad
&
{\sf b}  \doteq ~ ^3\mathrm{P}_0\go ^1\!\mathrm{S}_0,
\quad
&
{\sf c}  \doteq ~ ^3\mathrm{S}_1\go ^3\!\mathrm{S}_1,
\\[-.1cm]
& &
\\[-.1cm]
\hspace{-.6cm}
{\sf d}  \doteq ~ ^3\mathrm{D}_1\go ^3\!\mathrm{S}_1,
\quad
&
{\sf e}  \doteq ~ ^1\mathrm{P}_1\go ^3\!\mathrm{S}_1,
\quad
&
{\sf f}  \doteq ~ ^3\mathrm{P}_1\go ^3\!\mathrm{S}_1,
\end{array}
\er
as one can always do in the CMF within the s-wave approximation ~\cite{Kr10a}.
Therefore, we  only show separate results for the parity-conserving (PC) and parity-violating (PV) parts of the decay rates, which contain, respectively, the  (${\sf a}+{\sf c}+{\sf d}$) and (${\sf b}+{\sf e}+{\sf f}$) contributions. 
From  Table~\ref{T2} it can be concluded that:
\begin{enumerate}
\item
Although it is clear that the inclusion of relativity sensibly affects the results, there is gross agreement between analogous nonrelativistic and relativistic calculations, both without, and with SRC.
\item
The SRC, while not crucial for some decay rates, can significantly reduce  others, both in the nonrelativistic and in the relativistic cases.
\item
The decay rates $\Gamma_{n}$ and $\Gamma_{p}$ are both higher in RA2 than in RA1.
\item
Relativity tends to make $\Gamma_n$ become larger and $\Gamma_p$ smaller, and this effect is more pronounced in the RA2 approach for the propagator. As a consequence the relativistic $n/p$ ratio becomes significantly larger than the nonrelativistic one. Therefore, the relativistic approach, specially RA2, helps to solve the longstanding puzzle on the $\Gamma_{n}/\Gamma_{p}$ ratio~\cite{Ba02,Ga03}. (See also  Table~\ref{T3}.)
\end{enumerate}

\begin{table*}[htb] 
\caption{Results for the nonmesonic decay rates $\Gamma_n$, $\Gamma_p$ and
$\Gamma_{nm}=\Gamma_n+\Gamma_p$, and the ratio $\Gamma_{n}/\Gamma_{p}$ in
${}^{12}_{\Lambda}\textrm{C}$ for several OME models, namely, the $\pi$
and $\pi+K$ exchanges without and with SRC.
The present calculations with the approach RA2 for the propagators are compared  with previous relativistic calculations performed by
Refs.~\cite{Ra92,Co09,Co09a} and with the experimental data~\cite{Ki09,Ki06,Sa05,Ok05,Ag14}.
All results are in units of $\Gamma_{0}=2.50\times 10^{-12}$ MeV.
}\label{T3}
\begin{tabular*}{\textwidth}{p{3cm}@{\extracolsep{\fill}}|cccc}
\hline \hline\\[-.3cm]
Model 
&  $ \Gamma_{n}  $  & $ \Gamma_{p}  $  & $ \Gamma_{n}/\Gamma_{p} $  & $ \Gamma_{nm} $
\\ \hline
\underline{$\pi$ }\\
Present (RA2)          & $0.39$       &  $1.24$   & $0.31$ & $1.63$ \\
Ref.~\cite{Ra92}       & $0.27$       &  $1.32$   & $0.20$ & $1.62$
\\[0.06cm]
Ref.~\cite{Co09,Co09a} & $0.89$  &  $2.08 $  & $ 0.41  $  &  $ 2.95  $
\\[0.06cm]
\hline
\underline{$\pi$+SRC }\\
Present (RA2)&
                   $ 0.33$  &  $0.98 $  & $ 0.34$  &  $1.31$
\\[0.06cm]
Ref.~\cite{Ra92} & $0.06 $  &  $0.29$   & $ 0.20$  &  $0.35$
\\[0.06cm]
Ref.~\cite{Co09,Co09a} & $1.80$  &  $0.62 $  & $ 0.34  $  &  $ 2.42  $
\\[0.06cm]
\hline
\underline{$\pi+K$ }\\
Present (RA2)          & $0.41$  &  $ 0.98$  & $ 0.41  $  &  $ 1.39  $
\\[0.06cm]
Ref.~\cite{Co09,Co09a} & $1.24$  &  $1.59$     & $0.78$ &  $2.84$
\\[0.06cm]
\hline
\underline{$\pi+K$+SRC }\\
Present (RA2)& $0.33$  &  $ 0.82$  &  $ 0.40$  &  $ 1.15$
\\[0.06cm]
Ref.~\cite{Ra92} & $ 0.05 $  &  $0.36 $  & $ 0.15  $  &  $ 0.41  $
\\[0.06cm]
Ref.~\cite{Co09,Co09a} &
$0.96 $   &  $1.42 $   & $0.67$   &  $2.38$
\\[0.06cm]
\hline\hline
\underline{Experiment}
\\[0.06cm]
Ref.~\cite{Ki09}   &$0.23\pm 0.08 $ &$0.45\pm 0.10 $&  $- $  &  $- $
\\[0.06cm]
Ref.~\cite{Ki06}   &$- $ &$- $&  $0.51\pm 0.13\pm 0.05 $  &  $- $
\\[0.06cm]
Ref.~\cite{Sa05}   &$- $ &$- $& $-$ & $0.828\pm0.056\pm0.066$
\\[0.06cm]
Ref.~\cite{Ok05}  &$- $ &$- $&  $- $  &  $0.953\pm 0.032 $
\\[0.06cm]
Ref.~\cite{Ag14}   &$- $ &$0.65\pm 0.19$&  $- $  &  $- $
\\[0.06cm]
\hline\hline
\end{tabular*}
\end{table*}
%

Then, in Table \ref{T3}, are compared the present RA2 calculations for ${}^{12}_{\Lambda}\textrm{C}$
with previous relativistic calculations performed by Ramos \etal~\cite{Ra92}, and by Conti \etal~\cite{Co09,Co09a} for several OME models. The differences between the three theoretical calculations are very large.
We do not know the reason for such huge differences, although there are several possibilities. First, they could be due to the way in which the spectroscopic factors are evaluated, and how the states of the two emitted particles are antisymmetrized and normalized. Secondly, differences can arise simply from nume\-rical errors. To avoid those, we have checked step by step the entire relativistic calculation with the nonrelativistic one. 

In the same Table \ref{T3} are shown the pertinent experimental data produced by the KEK and FINUDA groups~\cite{Ki09,Ki06,Sa05,Ok05,Ag14}. 
When confronted with the several relativistic theoretical results, it is easy to discover that only the present eva\-luation within the $\pi+K+\mathrm{SRC}$ model agrees well with the available data. In particular, the good agreement for the ratio $\Gamma_{n}/\Gamma_{p}$  should be highlighted. 
The only significant discrepancy is with the experimental values for $\Gamma_n$ and $\Gamma_p$ obtained by KEK in Ref.~\cite{Ki09}. However, there is agreement with the experimental value for $\Gamma_p$ obtained by FINUDA in Ref.~\cite{Ag14}. As to the last column in Tables \ref{T2} and \ref{T3}, it is important to remark that, while all the listed calculations include only one-nucleon-induced transitions, the experimental values include also eventual two-nucleon-induced contributions.

\section{Summary and Final Remarks}
\label{Sec4}

Starting from the Fermi Golden Rule \rf{2.1}
we present   in Section \ref{Sec2} a relativistic formalism   to describe the nonmesonic weak decay of single-$\Lambda$ hypernuclei within the framework of the IPSM, with the dynamics represented by the $(\pi,K)$ OME model.
First, in Subsection \ref{Sec2A} we do this for hypernuclei whose cores have only closed subshells, and when the recoil effect is disregarded. Here, the  Dirac plane waves are expanded in 
spherical partial-waves, the multipole expansion of the propagator is done, and the two-body matrix element is properly antisymmetrized with regard to the two outgoing nucleons. Ma\-king use of the orthogonality condition \rf{2.16a} and explo\-ring the energy conserving $\delta$-function, the six momentum-space  integrals in Eq.\rf{2.1} are reduced to one in Eq.\rf{2.39}, to be performed numerically.
Next, the derived result is gene\-ralized to include hypernuclei with open-shell cores.
This is done by means of the spectroscopic factors given by Eq.\rf{2.41}, which are evaluated in second quantization, without recurring to the 
c.f.p.\ technique.
In this way we arrive at Eq.\rf{2.42}. 
Finally, in Subsection \ref{Sec2C} we discuss the recoil effect, which is important not only for the evaluation of angular distributions of the pairs of emitted nucleons, but also for the study of single kinetic energy spectra in light and medium-weight hypernuclei. 

Numerical results for ${}^{12}_{\Lambda}\textrm{C}$ are presented in Section \ref{Sec3}. Firstly, Table \ref{T2} shows the comparison between analogous nonrelativistic and relativistic calculations of  the transition rates $\Gamma_{n}$ and $\Gamma_{p}$. The parity-conserving and parity-violating contributions are given separately.
Such a comparison is crucial so that one can rely on relativistic calculations. 
Ho\-wever, it had never been done before.
The agreement is satisfactory, and the only difference worth mentioning is that the ratio  $\Gamma_{n}/\Gamma_{p}$ is appreciably higher in the relativistic calculation, agreeing better with experiment than the nonrelativistic one, specially when the RA2 approach to the propagators is used. 
Secondly, the present calculation is compared with two similar studies in Table \ref{T3}, from where it is clear that the discrepancies are very large.
Although there are important differences with our formalism, we could not find any justification to explain this. Done in the same table, the comparison of our calculation with the available experimental data is encouraging and favors the $\pi+K+\mathrm{SRC}$ OME model.

In short, we have achieved our goal of developing a reliable relativistic model for calculating the nonmesonic weak decay of $\Lambda$-hypernuclei, which can now be extended for similar weak decays in charmed nuclei. 
In parallel with this development, we intend to incorporate in our formalism the final state distortions of the outgoing nucleon waves induced by the interaction with the residual nucleus by making use of a relativistic optical potential.

\vspace{-.4cm}
\begin{acknowledgements}
Work partially supported by the Argentinean agencies Consejo Nacional de Investigaciones 
Cient\'ificas y T\'ecnicas - CONICET, Grant No.PIP 0377 (F.K.), and Fondo para la Investigaci\'on Cient\'ifica y Tecn\'ologica - FONCYT, Grant No.PICT-2010-2680 (F.K.), as well as by the Brazilian agencies Funda\c{c}\~ao de Amparo \`a Pesquisa do Estado de S\~ao Paulo - FAPESP,  Grants 2013/01790-5 (F.K.) and No. 2013/01907-0 (G.K.), and Conselho Nacional de Desenvolvimento Cient\'{\i}fico e Tecnol\'ogico - CNPq, Grant No.305894/2009-9 (G.K.). The work of C.E.F was supported by a post-graduate scho\-larship from Universidade Estadual Paulista. 
\end{acknowledgements}

\appendix

\section{Derivation of Eq. (\ref{2.16b})}
\label{A}
From the definition \rf{2.12b} it follows that
\br
&&\hspace{-.5cm}2\int\sum_{sm} d\ph\hat{j}^{-2}\delta_{jj'}
\ov{\ph s}{\kappa m}^\ast \ov{\ph s}{\kappa' m} \cdots 
\nn\\&&\hspace{-.5cm}
=2\hat{j}^{-2}\delta_{jj'}
(4\pi)^2\int d\ph\sum_{sm}i^{l-l'}
\sum_{\mu}(l\mu\fot s|jm)Y^*_{l\mu}(\ph)
\nn\\&&\hspace{-.5cm}
\times \sum_{\mu'}(l'\mu'\fot s|jm)Y_{l'\mu'}(\ph) \cdots 
\nn\\&&\hspace{-.5cm}
=2(4\pi)^{2} \hat{l}^{-2}
\delta_{jj'}\delta_{ll'}
\int d\ph\sum_{\mu}Y^*_{l\mu}(\ph)Y_{l\mu}(\ph) \cdots.
\label{A1}
\er
Now, if we use the relation
\br
4\pi \sum_{\mu}Y^*_{l\mu}(\ph)Y_{l\mu}(\ph)=\hat{l}^2,
\label{A2}
\er
we can solve the integral over the azimuthal angle to obtain
\br
&&2\int\sum_{sm} d\ph\hat{j}^{-2}\delta_{jj'}
\ov{\ph s}{\kappa m}^\ast \ov{\ph s}{\kappa' m} \cdots 
\nn\\
&&
=\delta_{\kappa\kappa'}(4\pi)^2
\int_{-1}^1 d\cos\theta \cdots.
\label{A3}
\er

\section{Relativistic single-particle wave functions\label{B}}

The evaluation of the matrix elements of the NMWD is made in the context of the 
IPSM. This means that the $\Lambda$ wave functions are those generated by spherically 
symmetric mesonic mean fields. That is, in solving the Dirac equations for the
single-particle level of $\Lambda$, one must use the meson mean fields from the 
$^{12}{\rm C}$ nucleus. This is similar in spirit to the works of  
Ramos~\etal~\cite{Ra91,Ra92}, where single-particle bound-state wave functions 
are obtained by solving the Dirac equation with static Lorentz-scalar and 
-vector Woods-Saxon potentials. 

The radial bound-state wave functions $F_\kappa(r)$ and $G_\kappa(r)$ in \rf{2.17} and 
corresponding energy eigenvalues $\varepsilon_\kappa$ for a single-particle state $\kappa$ 
for the $N$ or $\Lambda$ are obtained by solving the following Dirac equations:
\begin{eqnarray}
&&\hspace{-1.0cm}\left(\frac{d}{dr} + \frac{\kappa}{r}\right) F_\kappa
+ \left( \varepsilon_\kappa - V + S \right) G_\kappa =0,
\nn \\
&&\hspace{-1.0cm}\left(\frac{d}{dr} - \frac{\kappa}{r}\right) G_\kappa
- \left( \varepsilon_\kappa - V - S \right) F_\kappa =0,
\label{B1}
\end{eqnarray}
where the scalar potential $S = S(r)$  is
\begin{eqnarray}
S(r) = \Mass + g_\sigma \, \sigma(r),
\label{B2}
\end{eqnarray}
with $\Mass=\Mass_N$ and $g_\sigma = g^N_\sigma$ for the $N$, and $\Mass=\Mass_\Lambda$ 
and $g_\sigma=g^\Lambda_\sigma$ for the $\Lambda$; the vector potential $V=V(r)$ for the nucleon
is given by
\begin{eqnarray}
\hspace{-0.5cm}
V(r) = g^N_\omega \, \omega_0(r) + t_\kappa g_\rho \, \rho_0(r) + (t_\kappa + 1/2) e \, A_0(r),
\label{B3}
\end{eqnarray}
with $t_\kappa=1/2$ for the proton, $t_\kappa=-1/2$ for the neutron, and for the $\Lambda$ it is given~by
\begin{eqnarray}
V(r) =  g^\Lambda_\omega \, \omega_0(r).
\label{B4}
\end{eqnarray}
The meson and Coulomb fields satisfy the following Klein-Gordon and Poisson equations
\br
&&\hspace{-0.5cm}
\left(-\nabla^2 + m^2_\sigma\right) \sigma = - g^N_\sigma \, \rho^N_s 
- g_2 \sigma^2 - g_3 \sigma^3, \nn \\
&&\hspace{-0.5cm}
\left(-\nabla^2 + m^2_\omega\right) \omega_0 =  g^N_\omega \, \rho^N_B ,
\nn \\
&&\hspace{-0.5cm}
\left(-\nabla^2 + m^2_\rho\right) \rho_0  = 1/2 \, g_\rho \, \rho_3, 
\nn \\
&&\hspace{-0.5cm}
- \nabla^2 A_0 = e \, \rho_p ,
\er
with the densities given by
\br
\rho^N_s &=& \sum_{\kappa} \frac{n^N_\kappa}{4\pi r^2}  
\left(|F_\kappa|^2 - |G_\kappa|^2\right),
\nn \\
\rho^N_B &=& \sum_{\kappa} \frac{n^N_\kappa}{4\pi r^2} 
\left(|F_\kappa|^2 + |G_\kappa|^2\right),
\nn \\
\rho_3 &=& \sum_{\kappa} \frac{(-)^{t_\kappa-1/2}\,n^N_\kappa}{4\pi r^2} 
\left(|F_\kappa|^2 + |G_\kappa|^2\right),
\nn \\
\rho_p &=& \sum_{\kappa} \frac{(t_\kappa+1/2) \, n^N_\kappa}{4\pi r^2} 
\left(|F_\kappa|^2 + |G_\kappa|^2\right),
\er
where $n^N_\kappa$ are the nucleon occupancies of the state~$\kappa$. 

The system of equations is solved by iteration following the scheme of Ref.~\cite{Ho81}: 
(i)~we solve the Dirac equations for given initial ans\"atze for the $S$ and $V$ potentials; 
(ii)~the solutions for $F(r)$ and $G(r)$ are then used to solve the Klein-Gordon and Poisson 
equations and construct new potentials; and (iii) we put these into the Dirac equations and 
cycle until convergence to a prescribed precision is attained. Note that the nonlinear terms 
for the $\sigma$ field are put together with the scalar density $\rho^N_s$ in the iteration 
procedure.

The numerical values of the meson-nucleon parameters are those of the column 
NL3 \cite{La97} of Table~I in Ref.~\cite{vanGiai}, and for the meson-lambda couplings are those from Ref.~\cite{rufa}
(masses are given in~MeV):
\br
&& g^N_\sigma = 10.2169,\; g^N_\omega = 12.8675,  \; g_\rho = 8.9488, 
 \nn \\
&&e^2/4\pi = 1/137, \; g^\Lambda_\sigma = 0.464 \, g^N_\sigma, \; 
g^\Lambda_\omega = 0.481 \, g^N_\omega, \nn\\
&& g_2 = -10.4307~{\rm fm}^{-1},     \; g_3 = - 28.8851, \nn \\
&& m_\sigma = 508.1941, \; m_\omega = 782.501, \; m_\rho = 763.000, \nn \\
&& M_N = 939, \; M_\Lambda = 1116.06.  
\label{couplings}
\er

\begin{table}[h]
\caption{Single-particle energies for $^{12}{\rm C}$ and $^{12}_\Lambda{\rm C}$. (See text.) 
Experimental values for $^{12}{\rm C}$ are taken from Ref.~\cite{Ra92}, and for $^{12}_\Lambda{\rm C}$ 
from Ref.~\cite{Ch79}. All values are in MeV.}
\begin{ruledtabular}
\begin{tabular}{l|lll} 
                       &\hspace{-0.4cm}   Calculated  &\hspace{-0.6cm}  Experiment    \\
\hline\\[-.2cm]
p 1s$_{1/2}$            & -38.53 & -34    \\[0.1cm]
p 1p$_{3/2}$            & -13.52 & -15.96 \\[0.1cm]
n 1s$_{1/2}$            & -42.03 & -37    \\[0.1cm]
n 1p$_{3/2}$            & -16.65 & -18.72 \\[0.1cm]
$\Lambda$ \!1s$_{1/2}$  & -11.59 & -10.79  \\[0.1cm]
\end{tabular}
\end{ruledtabular}
\label{tab:sizes}
\end{table}

In Table~\ref{tab:sizes}, we present the single-particle energies 
for~$^{12}{\rm C}$ and $^{12}_\Lambda{\rm C}$. Note that these results are obtained without adjusting any parameters 
to fit experimental numbers. Clearly, a reasonable description of the experimental single-particle 
energies is achieved. Of course, a better description could be obtained by fine tuning the parameters,
but for the purposes of the present paper such a refinement is not necessary.

\vspace{0.5cm}

\section{Derivation of Eq.~(\ref{2.44})\label{C}}

Here we demonstrate the result \rf{2.44} starting from the definition \rf{2.20}  for
$S^\pi(p_1t_1,p_2t_2)$, \ie
\br
S^\pi(p_1t_1,p_2t_2)&\equiv&
\underset{s_1s_2}{\underset{m_\Lambda m_N}{\sum}}\int d\ph_1 
d\ph_2 \delta(\Delta_{j_N}-T_1-T_2-T_R)\nn\\&\times&
|\M^\pi(\pb_1\pb_2{s_1}{s_2}{t_1}{t_2}j_N m_Nj_\Lambda m_\Lambda)|^2 .
\er
Using the expansion \rf{2.18} and making the change of variable 
$\ph_2\to\ph_{12}$ as explained in Eq.~\rf{change}, we are free to perform the $\ph_1$ integration according to Eq.~\rf{2.16a}, and are left with
\br
&&S^\pi(p_1t_1,p_2t_2) \;=\;
\nn\\&&
(4\pi)^2\underset{\kappa_1 m_1s_2}
{\underset{m_\Lambda m_N}{\sum}}\int d\ph_{12} 
\delta(\Delta_{j_N}-T_1-T_2-T_R)
\nn\\&&\times 
\Big{|}\sum_{\kappa_2 m_2}\ov{\ph_{12} s_2}{\kappa_2 m_2}
\nn\\&&\times
\bra{p_1\kappa_1 m_1t_1p_2\kappa_2m_2t_2}{\sf \Delta}^\pi
\ket{\kappa_\Lambda m_\Lambda\kappa_N m_N}\Big{|}^2 .
\er
Then we  do angular momentum algebra as in \rf{2.22},
\br
&&S^\pi(p_1t_1,p_2t_2) \;=\;
\nn\\&&
(4\pi)^2\underset{\kappa_1 m_1s_2}{\underset{m_\Lambda m_N}{\sum}}
\int d\ph_{12}\delta(\Delta_{j_N}-T_1-T_2-T_R)
\nn\\&&\x
\underset{JM}{\underset{\kappa_2 m_2}{\sum}}
\ov{\ph_{12} s_2}{\kappa_2 m_2}^\ast 
\bra{p_1\kappa_1t_1 p_2\kappa_2t_2J}{\sf \Delta}^\pi
\ket{\kappa_\Lambda \kappa_NJ}^\ast 
\nn\\&&\x
(j_1m_1j_2m_2|JM)(j_\Lambda m_\Lambda j_N m_N|JM)
\nn\\&&\x
\underset{J'M'}{\underset{\kappa'_2 m'_2}{\sum}}
\ov{\ph_{12} s_2}{\kappa'_2 m'_2}
\bra{p_1\kappa_1t_1 p_2\kappa'_2t_2J'}{\sf \Delta}^\pi
\ket{\kappa_\Lambda \kappa_NJ'}
\nn\\&&\x
(j_1m_1j'_2m'_2|J'M')(j_\Lambda m_\Lambda j_N m_N|J'M')\,,
\er
to obtain
\br
&&S^\pi(p_1t_1,p_2t_2) \;=\;
\nn\\&&
(4\pi)^2\sum_{s_2}\int d\ph_{12}
\delta(\Delta_{j_N}-T_1-T_2-T_R)
\nn\\&&\x
\underset{\kappa'_2 m'_2}{\underset{\kappa_2 m_2}{\sum}}
\ov{\ph_{12} s_2}{\kappa_2 m_2}^\ast \ov{\ph_{12} s_2}{\kappa'_2 m'_2}
\nn\\&&\x
\underset{m_1M}{\underset{\kappa_1J }{\sum}}
(j_1m_1j_2m_2|JM)(j_1m_1j'_2m'_2|JM)
\nn\\&&\x
\bra{p_1\kappa_1t_1 p_2\kappa_2t_2J}{\sf \Delta}^\pi
\ket{\kappa_\Lambda \kappa_NJ}^\ast 
\nn\\&&\x
\bra{p_1\kappa_1t_1 p_2\kappa'_2t_2J}{\sf \Delta}^\pi
\ket{\kappa_\Lambda \kappa_NJ}.
\er
Due to the relation
\br
&&\sum_{ m_1M}(j_1m_1j_2m_2|JM)(j_1m_1j'_2m'_2|JM) \;=\;
\nn\\&&
\frac{\hat{J}^2}{\hat{j}_{2}^2}\delta_{m_2m'_2}\delta_{j_2j'_2}\,,
\er
this reduces to
\br
&&S^\pi(p_1t_1,p_2t_2) \;=\;
\nn\\&&
(4\pi)^2 
\underset{\kappa'_2s_2}{\underset{\kappa_2 m_2 }{\sum}}
\frac{\delta_{j_2j'_2}}{\hat{j}_{2}^2}
\int d\ph_{12}
\delta(\Delta_{j_N}-T_1-T_2-T_R)
\nn\\&&\x
\ov{\ph_{12} s_2}{\kappa_2 m_2}^\ast \ov{\ph_{12} s_2}{\kappa'_2 m_2}
\nn\\&&\x
\sum_{\kappa_1J}{\hat{J}^2}
\bra{p_1\kappa_1t_1 p_2\kappa_2t_2J}{\sf \Delta}^\pi
\ket{\kappa_\Lambda \kappa_NJ}^\ast 
\nn\\&&\x
\bra{p_1\kappa_1t_1 p_2\kappa'_2t_2J}{\sf \Delta}^\pi
\ket{\kappa_\Lambda \kappa_NJ}.
\er
Finally, using (\ref{2.16b}) one gets \rf{2.44}.



\begin{thebibliography}{99}
%
\bibitem{Bo12} E.\ Botta, T.\ Bressani, and G.\ Garbarino, 
Eur.\ Phys.\ J.\ A \textbf{48}, 41 (2012).
%
\bibitem{Feliciello:2014ola} 
  A.~Feliciello,
  Few Body Syst.\  {\bf 55}, 605 (2014).
%
\bibitem{Garbarino:2013rwa} 
  G.~Garbarino,
  Nucl.\ Phys.\ A {\bf 914}, 170 (2013).
%
\bibitem{Bufalino:2013qwa} 
  S.~Bufalino,
  Nucl.\ Phys.\ A {\bf 914}, 160 (2013).
%
\bibitem{OsetRamos} E. Oset and A. Ramos, Prog. Part. Nucl. Phys. {\bf 41}, 191 (1998).
%
\bibitem{AlbericoGarbarino} W. M. Alberico and G. Garbarino, Phys. Rep. {\bf 369}, 1 (2002) 1.
%
\bibitem{Parreno} A. Parre\~no, Lec. Notes Phys. {\bf 724}, 141 (2007).
%
\bibitem{Outa} H. Outa, in: Hadron Physics, IOS Press, Amsterdam, 2005, p. 219.
%
\bibitem{Pocho} J. Pochodzalla, Acta Phys. Pol. B {\bf 42}, 833 (2011).
%
\bibitem{PDG} K.A. Olive et al. (Particle Data Group), 
Chin. Phys. C {\bf 38}, 090001 (2014).
%
\bibitem{Du96} J. F. Dubach, G. B. Feldman, B. R. Holstein and L. de la Torre,
Ann. Phys. (N.Y.) {\bf 249}, 146 (1996).
%
\bibitem{Pa97} A. Parre\~{n}o, A. Ramos, and C. Bennhold, 
Phys. Rev. C \textbf{56}, 339 (1997).
%
\bibitem{It02} K. Itonaga, T. Ueda, and  T. Motoba, 
Phys. Rev. C \textbf{65}, 034617 (2002).
%
\bibitem{It03} K. Itonaga,  T. Motoba,  and T. Ueda, 
Mod. Phys. Lett. A \textbf{18}, 135 (2003).
%
\bibitem{It08} K. Itonaga,  T. Motoba,  T. Ueda, and Th.A. Rijken, 
Phys. Rev. C \textbf{77}, 044605 (2008).
%
\bibitem{Ba02} C. Barbero, D. Horvat, F. Krmpoti\'{c}, T. T. S. Kuo, Z. Naran\v{c}i\'c, 
and D. Tadi\'{c}, Phys. Rev. C \textbf{66}, 055209 (2002).
%
\bibitem{Kr03} F. Krmpoti\'c and  D. Tadi\'c, Braz. J. Phys. {\bf 33}, 187 (2003).
%
\bibitem{Ba03} C. Barbero, C. De Conti, A. P. Gale\~ao, and F. Krmpoti\'c, 
Nucl. Phys. \textbf{A726}, 267 (2003).
%
\bibitem{Ba05}
C. Barbero, A. P. Gale\~ao, and F. Krmpoti\'c, 
Phys. Rev. C \textbf{72},  035210  (2005).
%
\bibitem{Ba07} C. Barbero, A. P. Gale\~ao, and F. Krmpoti\'c, 
Phys. Rev. C \textbf{76},  054321  (2007).
%
\bibitem{Ba08} C. Barbero, A. P. Gale\~ao, M. S. Hussein, and  F. Krmpoti\'c,
 Phys. Rev. C\textbf{78},  044312 (2008)
%
\bibitem{Ba10} E. Bauer, A. P. Gale\~ao, M. S. Hussein, and F. Krmpoti\'c,
Nucl. Phys. A\textbf{834}, 599c (2010).
%
\bibitem{Kr10} F. Krmpoti\'c,  A. P. Gale\~ao, and M.S. Hussein,
AIP Conf. Proc. {\bf 1245},  51 (2010).
%
\bibitem{Kr10a} F. Krmpoti\'c, Phys. Rev. C \textbf{82}, 05520 (2010).
%
%
\bibitem{Kr14}  F. Krmpoti\'c,  Few Body Syst. \textbf{55}, 219 (2014).
%
\bibitem{Kr14a} F. Krmpoti\'c and C. De Conti,
Int. J. of Mod. Phys. E {\bf 23}, 1450089 (2014).
%
\bibitem{Go11} I. Gonzalez, C. Barbero, A. Deppman, S. Duarte, F.
Krmpoti\'c, and O. Rodriguez, J. Phys. G: Nucl. Part. Phys. \textbf{38},
115105 (2011).
%
\bibitem{Tyap} A. A. Tyapkin, Yad. Fiz. {\bf 22}, 181 (1975).
%
\bibitem{Iwao} S. Iwao, Lett. Nuovo Cim. {\bf 19}, 647 (1977).
%
\bibitem{Do77}  C.B. Dover and S.H. Kahana, Phys. Rev. Lett. \textbf{39}, 1506 (1977).
%
\bibitem{GatPac} R. Gatto and F. Paccanoni, Nuovo Cim. A {\bf 46}, 313 (1978).
%
\bibitem{Kol} 
  N.~N.~Kolesnikov, D.~I.~Zhukovitsky, V.~A.~Kopylov, and V.~I.~Tarasov,
  Sov.\ J.\ Nucl.\ Phys.\  {\bf 34}, 533 (1981)
  [Yad.\ Fiz.\  {\bf 34}, 957 (1981)].
%
\bibitem{Bham} G. Bhamathi, Phys. Rev. C {\bf 24}, 1816 (1981).
%
\bibitem{Bando} H. Bando and M. Bando, Phys. Lett. B {\bf 109}, 164 (1982).
%
\bibitem{Gib} B. F. Gibson, C. B. Dover, G. Bhamathi, and D. R. Lehman, 
Phys. Rev. C {\bf 27}, 2085 (1983).
%
\bibitem{StaTsa} N. I. Starkov and V. A. Tsarev, Nucl. Phys. {\bf A450}, 507 (1986).
%
\bibitem{Cai} C. H. Cai, L. Li, Y. H. Tan, P. Z. Ning, Europhys. Lett. {\bf 64}, 448 (2003).
%
\bibitem{Tsu1} K. Tsushima and F. C. Khanna, Phys. Lett. B {\bf 552}, 138 (2003).
%
\bibitem{Tsu2} K. Tsushima and F. C. Khanna, Phys. Rev. C {\bf 67}, 015211 (2003).
%
\bibitem{Tsu3} K. Tsushima and F. C. Khanna, J. Phys. G {\bf 30}, 1765 (2004).
%
\bibitem{Bu92} S. A. Bunyatov, V. V. Lyukov, N. I. Starkov, and V. A. Tsarev, 
Sov. J. Part. Nucl. \textbf{23}, 253 (1992).
%
\bibitem{exp1} Yu. Batusov, S. A. Bunyatov, V. V. Lyukov, V. M. Sidorov, A. A. Tyapkin,
and V. A. Yarba, JETP Lett. {\bf 33}, 56 (1981) [Pis'ma Zh. Eksp. Teor. Fiz. {\bf 33},
56 (1981)].
%
\bibitem{exp2} V. V. Lyukov, Nuovo Cim. A {\bf 102}, 583 (1989).
%
\bibitem{BroWei} R. Brockmann and W. Weise, Phys. Lett. B {\bf 69}, 167 (1977).
%
\bibitem{Ra91} A. Ramos, C. Bennhold, E. van Meijgaard, and B.K. Jennings, 
Phys. Lett. B \textbf{264}, 233 (1991).
%
\bibitem{Ra92} A. Ramos, E. van Meijgaard, C. Bennhold, and B.K. Jennings, 
Nucl. Phys. \textbf{A544}, 703 (1992).
%
\bibitem{Co09} F. Conti, ``A relativistic model for the non-mesonic weak decay of the 12C hypernucleus'', PhD Thesis, University of Pavia, Italy, November 2009.
%
\bibitem{Co09a} 
  F.~Conti, A.~Meucci, C.~Giusti and F.~D.~Pacati,
  arXiv:0912.3630 [nucl-th].
%
\bibitem{Ri15} P. Ring, private communication.
%
\bibitem{Hagino} K. Hagino and J. M. Yao, in ``Relativistic Density Functional for Nuclear Structure" (World Scientific, Singapore, 2015). 
%
\bibitem{Na72} R. Almar, O. Civitarese, F. Krmpoti\'c, and J. Navaza, 
Phys. Rev. C \textbf{6}, 187 (1972); J. Navaza, 
``Descripci\'on de N\'ucleos Vibracionales con el Modelo Unificado mediante T\'ecnicas Diagram\'aticas'', 
University of La Plata, Argentina, 1972.
%
\bibitem{de63} A. de-Shalit and I. Talmi, {\em Nuclear Shell Theory} (Academic Press, New York, 1963).
%
\bibitem{Ki09} M. Kim,~\etal, Phys. Rev.  Lett. \textbf{103}, 182502 (2009).
%
\bibitem{Ki06} M. J. Kim ~\etal, Phys. Lett. B \textbf{641}, 28 (2006).
%
\bibitem{Sa05} Y. Sato \etal, Phys. Rev. C \textbf{71}, 025203 (2005).
%
\bibitem{Ok05} S. Okada~\etal, Nucl. Phys. \textbf{A754}, 178c (2005).
%
\bibitem{Ag14} M. Agnello~\etal, Phys. Lett.  B \textbf{738}, 499 (2014).
%
\bibitem{Ga13} A.P. Gale\~ao, C. Barbero, C. De Conti, and F. Krmpoti\'c, 
AIP Conf.\ Proc.\ \textbf{1529}, 247 (2013).
%
\bibitem{Co14} C. De Conti, C. Barbero, A. P.  Gale\~ao, and F. Krmpoti\'c, 
AIP Conf.\ Proc.\ \textbf{1625}, 181 (2014).
%
\bibitem{Se86} B.~D.~Serot and J.~D.~Walecka, 
Adv.\ Nucl.\ Phys.\  {\bf 16}, 1 (1986).
%
\bibitem{Ma89} P.M.M. Maessen, Th.A. Rijken, and J.J. de Swart, 
Phys. Rev. C \textbf{40}, 226 (1989).
%
\bibitem{Do85}  M. Doi, T. Kotani, and E. Takasugi, 
Prog. Theor. Phys. Supplement \textbf{83}, 1 (1985).
%
\bibitem{Ho81} C.J. Horowitz and B.D. Serot, Nucl. Phys. \textbf{A368}, 503 (1981).
%
\bibitem {Ba01} C. Barbero, D. Horvat, F. Krmpoti\'{c}, Z. Naran\v{c}i\'c, and D. Tadi\'{c}, 
Fizika \textbf{B 10}, 1 (2001).
%
\bibitem{Sas05} K. Sasaki, M. Izaki, and  M. Oka, Phys. Rev. C \textbf{71}, 035502 (2005).
%
\bibitem{Bl63} M.M. Block and R.H. Dalitz, Phys. Rev. Lett. \textbf{11}, 96  (1963).
%
\bibitem{Ga03}  G. Garbarino, A. Parre\~{n}o, and A. Ramos, Phys. Rev.  Lett. \textbf{91}, (2003) 112501;
Phys. Rev. C \textbf{69}, 054603 (2004).
%
\bibitem{La97} G. A. Lalazissis, J. K\"onig, and P. Ring, 
Phys. Rev. C \textbf{55}, 540 (1997).
%
\bibitem{vanGiai} Wenjui Long, Jie Meng, N. van Giai, and S.-G. Zhou, Phys. Rev. C {\bf 69}, 034319 (2004).
%
\bibitem{rufa} M. Rufa, J. Schaffner, J. Maruhn, H. St\"ocker, W. Greiner, and
P.G. Reinhard,  Phys. Rev. C {\bf 42}, 2469 (1990).
%
\bibitem{Ch79}
R.E. Chrien \etal, Phys. Lett. B \textbf{89}, 31 (1979).
%
\end{thebibliography}
\end{document}